\documentclass[prd,superscriptaddress,nofootinbib,11pt, tightenlines]{revtex4-2} 
\usepackage{epsfig}  
\usepackage{amsmath} 
\usepackage{amsfonts}  
\usepackage{xfrac}
\usepackage{float}    
\usepackage{amssymb}    
\usepackage{slashed} 
\usepackage{color}       
\usepackage{pbox}
\usepackage{subfigure}
\usepackage{array,multirow,makecell}
\usepackage[colorlinks,citecolor=blue, linkcolor = blue, urlcolor = blue]{hyperref}
\usepackage{tabularx}
\usepackage{titlesec} 
\usepackage{scrextend}
\usepackage{natbib} 
\usepackage{array,multirow,makecell}
\usepackage{lipsum}
\usepackage[normalem]{ulem} %to strike the words
\usepackage{tikz,xcolor,hyperref}
\usepackage{marginnote}
\def \met  {\mbox{${E\!\!\!\!/_T}$}}
\begin{document}

\title{Implications of Sgr\,A$^\ast$ on the ${\gamma}$-rays searches of Bino Dark Matter with ${(g-2)_\mu}$}

\author{Utpal Chattopadhyay}
\email{tpuc@iacs.res.in}
\affiliation{School of Physical Sciences, Indian Association for the Cultivation of Science, Jadavpur, Kolkata 700032, India}

\author{Debottam Das}
\email{debottam@iopb.res.in}
\affiliation{Institute of Physics, Sachivalaya Marg, Bhubaneswar, 751005, India}
\affiliation{Homi Bhabha National Institute, Training School Complex, Anushakti Nagar, Mumbai 400094, India}

\author{Sujoy Poddar}
\email{sujoy.phy@gmail.com}
\affiliation{Department of Physics, Diamond Harbour Women’s University,
Diamond Harbour Road, Sarisha, South 24 Parganas, West Bengal — 743368, India}

\author{Rahul Puri}
\email{rahul.puri@iopb.res.in}
\affiliation{Institute of Physics, Sachivalaya Marg, Bhubaneswar, 751005, India}
\affiliation{Homi Bhabha National Institute, Training School Complex, Anushakti Nagar, Mumbai 400094, India}

\author{Abhijit Kumar Saha}
\email{abhijit.saha@iopb.res.in}
\affiliation{Institute of Physics, Sachivalaya Marg, Bhubaneswar, 751005, India}
\affiliation{Homi Bhabha National Institute, Training School Complex, Anushakti Nagar, Mumbai 400094, India}

\begin{abstract} 
{\bf Abstract:} 
We analyse the impact of dark matter density spike around the Milky Way's supermassive black hole (SMBH), Sgr\,A$^*$, in probing the Bino-dominated neutralino dark matter (DM)
$\tilde \chi_1^0$ within the MSSM, which typically produces relatively faint signals in the conventional DM halos. In particular, we explore the indirect search prospects of sub-TeV Bino-Higgsino and Bino-Wino-Higgsino DM in the MSSM, consistent with the supersymmetric predictions required to explain the anomalous magnetic moment of the muon. Typical over-abundance of Bino DM is ameliorated with slepton and/or Wino coannihilations. The lightest neutralino, thus, may be associated with a compressed supersymmetric particle spectrum, which, in general, is difficult to probe at conventional LHC searches. Similarly, for a rather tiny Higgsino mixing, $\tilde \chi_1^0$ does not offer much prospect to assess its predictions at dark matter direct detection searches. Accommodating the inclusive effects of density spike, here, we
present the requisite boost factor to facilitate $\gamma-$ray searches 
of Bino-dominated DM in the MSSM, especially focusing on the Fermi-LAT and HESS observations.
\end{abstract}

\maketitle

%%%%%%%%%%%%%%%%%%%%%%%%%%%%%
\section{Introduction}
%%%%%%%%%%%%%%%%%%%%%%%%%%%%%
For decades, it has been well recognized that the Minimal Supersymmetric Standard Model (MSSM)~\cite{Nilles:1983ge, Haber:1984rc, Wess:1992cp, Drees:2004jm, Martin:1997ns} offers a natural candidate for dark matter (DM)~\cite{Clowe:2006eq, Bertone:2004pz, Jungman:1995df} in the Universe if the $R$-parity is conserved \cite{Martin:1996kn}. The neutral supersymmetric (SUSY) partners of the Higgs (Higgsinos) and electroweak gauge bosons (gauginos) can be assembled into four neutralinos, $\tilde{\chi}_{1,2,3,4}^0$, with $\tilde{\chi}_1^0$ 
referring to the lightest among them. 
In MSSM, $\tilde\chi_1^0$ can be considered to be the lightest supersymmetric particle (LSP) to serve the role of a viable DM candidate \cite{Jungman:1995df, Bertone:2004pz}.
Depending on the individual compositions of the gauginos and the Higgsinos, the neutralino DM could become Bino, Wino, or Higgsino-dominated or any suitable admixtures of them.

On the experimental front, the LHC null observation of coloured sparticles has excluded the masses of squarks and gluinos smaller than a few TeV \cite{ATLAS:2020syg, CMS:2019zmd}. The uncoloured particles, such as elctroweakinos and sleptons, can still be in the sub-TeV range since LHC constraints are relatively weaker due to smaller production cross-sections \cite{Canepa:2019hph} (see, also \cite{Chakraborti:2015mra}). 
Importantly, lighter electroweakinos and sleptons are pivotal in reconciling the difference observed between the measured and the theoretically predicted values of the muon's anomalous magnetic moment, $a_\mu=(g-2)_\mu/2$. Specifically, at one-loop level, loops containing electroweakinos and lighter sleptons e.g.,  $\tilde{\chi}^{-}-\tilde{\nu}_\mu$ and $\tilde{\chi}^{0}-\tilde{\mu}$ contribute to $\delta a_\mu= a_\mu^{\rm Exp}-a_\mu^{\rm SM}$ \cite{Moroi:1995yh, Martin:2001st, Stockinger:2006zn, Athron:2015rva}.
The disagreement over $a_\mu$ value was previously reported by Brookhaven National Laboratory (BNL) E821 experiment \cite{Muong-2:2006rrc}. Combining with Fermilab Muon g-2 experiments \cite{Muong-2:2021ojo}, it establishes a deviation of $4.2\sigma$ from the value predicted by the SM. The recent Fermilab E989 experimental data from Run-2 and Run-3 have also confirmed the discrepancy, giving a new world average of $5.1\sigma$ \cite{Muong-2:2023cdq}. 

In the past, the MSSM parameter space analyses combining neutralino DM and $\delta a_\mu$ were thoroughly performed in several works\,\cite{Lopez:1993vi, Chattopadhyay:1995ae, Moroi:1995yh, Chattopadhyay:2000ws, Martin:2001st, Heinemeyer:2003dq, Stockinger:2006zn, Cho:2011rk, Endo:2013lva, Endo:2013bba, Chakraborti:2014gea, Chakraborti:2015mra, Chowdhury:2015rja,  Kowalska:2015zja, Gomez:2018efz, Chakraborti:2021mbr, Chakraborti:2021kkr, Ali:2021kxa, Lindner:2016bgg, Bisal:2023iip}. A dominantly Bino-like light $\tilde \chi_1^0$ along with co-annihilating sleptons (in particular, $\tilde{\mu}$, $\tilde{\nu}_\mu$) are favoured by $\delta a_\mu$, especially if the observed DM abundance has to come entirely from the lightest neutralino having masses
$\lesssim 500$ GeV. 
 To note, Bino-slepton coannihilation $\tilde\chi_1^0 \tilde{l} \xrightarrow{l} 
{l \gamma }$ does not need any tuning in the neutralino mixing matrix to produce the correct relic abundance. This may not be true for a significantly mixed $\tilde \chi_1^0$ in the form of Bino-Higgsino-like DM, which can also reconcile the observed DM abundance and the discrepancy in $\delta a_\mu$. In this case, apart from the fact that fine-tuning the different elements in the mixing matrices needs to be adhered to, the null results in the Spin-Independent (SI) dark matter direct detection 
experiments (DD) \cite{PandaX-II:2020oim, XENON:2023cxc, XENON:2018voc, LUX:2017ree, PandaX-II:2017hlx, XENON:2020kmp, PICO:2016kso, LUX:2016sci, PandaX-II:2016wea, LUX-ZEPLIN:2022xrq}
may severely restrict the 
MSSM parameter space \cite{Baer_2016, Badziak:2017the, Profumo:2017ntc,Abdughani:2019wai}. In fact, SI-DD searches
directly curb the Bino-Higgsino-Higgs coupling.
On the other hand, in the presence of a minimal Wino composition in $\tilde \chi_1^0$ (assuming the Higgsino fraction to be negligibly small), the SI-DD rates will never reach the sensitivity of the present or future experiments for a dominantly Bino-like DM. It follows from the fact that the Higgs coupling to the LSP pair is proportional to the product of their appropriate mixing matrix elements corresponding to the Higgsino and gaugino components \cite{Rosiek:1989sce, Martin:1997ns}. 

In recent times, Ref.~\cite{Abdughani:2019wai, Bisal:2023iip} explore the regions
of the allowed MSSM parameter space, compatible with the $a_\mu$ anomaly, DM relic density, DD limits, and the latest LHC Run-2 data. 
It turns out that two specific regions will be of interest
for future searches : (1)
a dominantly Bino-like light $\tilde \chi_1^0$ with minimal Higgsino contributions ($M_1\ll\mu$), and (2) a compressed scenario of Bino, Wino, a minimal Higgsino ($M_1\lesssim M_2 \ll\mu$) components. On both occasions, masses of sleptons are not far from that of the LSP. In the subsequent sections, we
refer them as $\tilde B_{\tilde H}$ and $\tilde B_{\tilde W\tilde H}$ 
zones of the MSSM parameter space.
The SI-DD search prospects of the DM can be improved to some extent if one incorporates all the electroweak vertex corrections in the computations of SI-DD of the $\tilde\chi_1^0$ DM  \cite{Bisal:2023fgb, Bisal:2023iip}. 
Even then, a substantial part of the parameter space, specifically with higher $\mu$ values, will be out of the periphery of future SI-DD searches. Our aim in this work would be to survey the $\tilde B_{\tilde H}$ and $\tilde B_{\tilde W\tilde H}$ regimes of the MSSM parameter space through the possible indirect searches 
of $\gamma$-ray observations, which otherwise may escape the LHC limits
of the direct production of the SUSY particles. Specifically, we will use Fermi-LAT and HESS observations for scrutinizing the parameter space \cite{Fermi-LAT:2015bhf, Aharonian:2009zk, HESS:2016mib, HESS:2016pst}. 

It is well-known that the $\gamma$-ray flux from the DM pair annihilations completely depends on the Wino or Higgsino components
of the LSP. Naturally, a high value for either one of the components or both is desirable for a potential rise in the $\gamma-$ray flux. However, the observed abundance of DM along with the compatibility with $\delta a_\mu$ and the SI-DD bounds strongly restrict the amount of Wino or Higgsino components in the LSP that belong to the $\tilde B_{\tilde H}$ and $\tilde B_{\tilde W\tilde H}$ regimes. This, in turn, may result in suppression of the $\gamma$-ray spectra from the DM-DM annihilations inside the conventional DM halos, thus posing a challenge to the standard indirect searches of the dark matter. The dark matter signal enhancement in the galactic halos is possible in the presence of a {\it boost factor}. Different proposals explaining the boost factor, such as Sommerfield enhancement \cite{Hisano:2004ds, Arkani-Hamed:2008hhe, Pospelov:2008jd}, Breit-Wigner resonance \cite{Ibe:2008ye} and enhancement in the local DM density, are prevalent in the literature. However, for a dominantly Bino-like DM, away from the resonance, the rise in the local DM density may become the only possibility to produce a much-needed boost in the $\gamma$-ray signals of the DM coming from the Galactic Center (GC) or any 
dwarf galaxies. 

A boost such as that mentioned above can potentially be realized through the DM density spikes around the supermassive black hole
(SMBH) at the center of the Milky Way  
\cite{Gondolo:1999ef}. Recently, the indirect evidence of a dark matter density spike around two stellar-mass black holes has been reported \cite{Chan:2022gqd}. One can find that the high density of DM in the form of a spike within the gravitational zone of SMBH can magnify the DM annihilation rates to some extent, resulting in a bright source to the $\gamma$-ray telescope. Earlier works \cite{Shelton:2015aqa, Shapiro:2016ypb, Chiang:2019zjj, Arina:2015zoa, Sandick:2016zeg, Cheng:2023hzw, Balaji:2023hmy, Christy:2023tdv, Cheng:2022esn, Alvarez:2020fyo, Cheng:2023dau, Shapiro:2022prq, Yang:2024jtp} in this direction have anticipated that such magnification in the DM annihilation signals can indeed offer a unique probe of the dark matter models. 
Considering Milky Way's SMBH Sgr A$^*$\,\cite{Balaji:2023hmy} at the backdrop of our analysis, we exemplify the DM density profile and subsequently calculate the observable $\gamma$-ray flux in the MSSM for both $\tilde B_{\tilde H}$ and $\tilde B_{\tilde W\tilde H}$ cases. The predictions obtained will further
be substantiated by the public results of Fermi-LAT\,\cite{Fermi-LAT:2015bhf} and HESS telescopes\,\cite{HESS:2016pst} that consider GC as source.

We organize our paper as follows. Sec.\,\ref{Sec:2} briefly discusses the properties of $\tilde \chi_1^0$. The dependence of $\gamma$-ray flux on the astrophysical and particle physics inputs has also been summarized. Here,
we detail the astrophysical model, which can lead to a spike
in the DM density distribution within the gravitational
influence zone of SMBH. In Sec.\,\ref{sec:3}, we summarize the precision observables, e.g., supersymmetric contributions to anomalous magnetic moments of muon ($a^{\rm SUSY}_\mu$) and other constraints like the limits from the SUSY searches at the collider experiments. The constraints from
the flavour physics and non-accelerator-based experiments are
also noted. Sec.\,\ref{sec:4} presents the numerical results of the $\gamma$-ray flux for
$\tilde B_{\tilde H}$ and $\tilde B_{\tilde W\tilde H}$ regimes in the MSSM. Finally, we conclude in 
Sec.\,\ref{sec:5}.

%%%%%%%%%%%%%%%%%%%%%%%%%%%%%%%%%%%%%%%%%
\section{Photon spectra of 
Neutralino dark matter in the presence of the Milky Way’s
supermassive black hole}
\label{Sec:2}
%%%%%%%%%%%%%%%%%%%%%%%%%%%%%%%%%%%%%%%%%
In the MSSM, the supersymmetric partner of the $U(1)_Y$ gauge boson, $\tilde {B}$, $SU(2)_L$ neutral gauge boson, $\tilde W^0$, mix with the supersymmetric partners of the two MSSM Higgs bosons, known as down and up type Higgsinos $\tilde{H}_d^0$ and $\tilde{H}_u^0$ respectively.
The mass matrix of the neutralino sector in the basis $(\tilde B, \tilde W^0, {\tilde H}^0_u, {\tilde H}^0_d)$ is given by:
\begin{align}  \mathbb{M}=\left(
    \begin{array}{c c c c}
        M_1 & 0 & -M_Zs_W c_\beta & M_Z s_W s_\beta \\[6pt]
        0 & M_2 & M_Zc_W c_\beta & -M_Z c_W s_\beta\\[6pt]
        -M_Zs_W c_\beta & M_Zc_W c_\beta & 0 & -\mu\\[6pt]
        M_Zs_W s_\beta & -M_Zc_W s_\beta & -\mu & 0
    \end{array}\right),
    \label{eq:mchi_mat}
\end{align}
where $M_1$ and $M_2$ are the Bino and Wino masses, respectively, $\mu$ is the Higgs mixing parameter,
$c_\beta=\cos\beta$, $s_\beta=\sin\beta$ with
$\tan\beta=\frac{v_2}{v_1}$, $v_2$ and $v_1$ being the vacuum expectation values of the two Higgs doublets ($H_u$ and $H_d$). Also, $c_W=\cos\theta_W$, $s_W=\sin\theta_W$ where $\theta_W$ is the weak mixing angle. 

The lightest neutralino, widely considered as 
a potential DM candidate in an R-parity conserving MSSM can be expressed as,
\begin{align}
\tilde{\chi}_1^0=N_{11}{\tilde B}+N_{12}{\tilde W^0}+N_{13}\tilde{H}^0_u+N_{14} {\tilde H}^0_d,
\end{align}
where $N$ is $4\times4$ unitary matrix that diagonalizes $\mathbb{M}$ in Eq.(\ref{eq:mchi_mat}). 
The LSP could be of different natures in different parts of the MSSM parameter space depending on the relative mass values of the gaugino and Higgsino mass parameters.
For example, if $M_1\ll M_2,\mu$, it turns out that $N_{11}\gg N_{1i}$  $( i=2$\,-\,$4)$ and the LSP becomes Bino-like. In another case, when $M_1\lesssim \mu\ll M_2$, the DM becomes Bino-Higgsino-like. It is also possible that $M_1\sim M_2\lesssim \mu$; in that case, the LSP will be Bino-Wino-Higgsino type. However, following the strong limits of the SI-DD searches, in both
 scenarios, $\mu \gg M_1$ is assumed. 
 
 We now begin with the prerequisites
to calculate photon flux in the DM halo. Subsequently, we will elaborate
on the astrophysical model, which may lead to a spike in the DM density distribution around an SMBH. 

%%%%%%%%%%%%%%%%%%%%%%%%%%%%%%%%%%%% 
 \begin{center}
 {\textbf{Photon Flux and Astrophysical Model}} 
 \end{center}
%%%%%%%%%%%%%%%%%%%%%%%%%%%%%%%%%%%%%
The indirect detection techniques are based on the detection of secondary particles that come out of dark matter (DM) annihilations in the over-dense region of the universe \cite{Gaskins:2016cha, Slatyer:2021qgc, Hooper:2018kfv}. The observable signals in the form of photons ($\gamma$), positrons ($e^+$), antiprotons ($\overline{p}$), neutrinos ($\nu$), etc may be searched for the quest of the DM sources. Neutral particles ($\gamma, \nu$, etc.) travel undisturbed through the interstellar medium and are more accurate in describing their origin. On the other hand, charged particles encounter several interactions, such as the galactic winds, interstellar magnetic field, etc, before reaching the observatory and thus are subject to cosmic-rays (CR) propagation models \cite{Donato:2003xg, Delahaye:2007fr, Salati:2021hnj, Slatyer:2017sev}. The differential photon flux originating from the DM-DM annihilations in the galactic halo can be calculated as \cite{Gaskins:2016cha, Slatyer:2021qgc}:
 
\begin{align}
    \dfrac{d\Phi}{dE_\gamma}&=\frac{1}{4\pi}\dfrac{\langle\sigma v \rangle}{2m_{\tilde\chi_1^0}^2}\dfrac{dN}{dE_\gamma}\int_{\Delta\Omega}d\Omega\int_{\rm LOS}d\ell\;\rho_{\tilde\chi_1^0}^2(r)\\
    &=\dfrac{\langle\sigma v \rangle}{2m_{\tilde\chi_1^0}^2}\dfrac{dN}{dE_\gamma}\times J_{\rm halo}[\rho_{\tilde\chi_1^0}(r)], \label{eq:ordFlux}
\end{align}
where $\langle\sigma v\rangle$ is the velocity averaged DM-DM annihilation cross-section with $v \sim 0.001$, 
$m_{\tilde\chi_1^0}$ is the mass of the DM, and ${dN}/{dE_\gamma}$ is the differential photon  spectrum per DM-DM annihilation. 
The latter can be obtained by summing the spectra of individual annihilation channels, each weighted by the corresponding branching ratio, such that,
\begin{align}
    \frac{dN}{dE_\gamma}=\sum_i {\rm Br(\tilde\chi_1^0\,\tilde\chi_1^0}\to f_i)\,\frac{dN_i}{dE_\gamma}, \label{eq:dndeIndiv}
\end{align}
where $f_i$ are the final state particles in the $i$-th annihilation channel.
The remaining term in Eq.(\ref{eq:ordFlux}), known as the $J-$factor, is of astrophysical origin and is a functional of the DM density profile. In some specific cases, it may also depend on particle physics inputs such as the DM mass and DM annihilation cross-section (see \textit{e.g.} \cite{Shapiro:2016ypb, Balaji:2023hmy}). The functional form of DM density profile $\rho(r)$ has many proposals based upon different considerations~\cite{1977ApJ...218..592G, Navarro:1995iw, 1965TrAlm...5...87E, Salucci:2000ps, Moore:1999gc,  Kravtsov:1997dp}. 
Around an SMBH ({\it e.g.} Sgr\,A$^*$), a spike is expected to appear due to DM's local overdensity within its gravitational influence zone. The radius of the gravitational influence of the
SMBH is $r_{\rm\bf h}=GM/v_0^2$. 
Here, $M$ is the mass of the
SMBH, $G$ is Newton's universal gravitational constant, and $v_0$ is the velocity dispersion of DM in the halo outside the spike.
 This, in turn, may enhance the detection possibility of a DM candidate, even with a suppressed annihilation rate. 
 Following the references \cite{Shelton:2015aqa, Johnson:2019hsm, Fields:2014pia}, the possible DM density spike can be modelled by various connected power laws with different parameters in varying regions around the SMBH.
  The multiple regions of the profile are described as follows :

 \begin{itemize}
     \item Outside the gravitational influence region of the SMBH, the DM halo can be characterised as a generalized NFW profile~\cite{Navarro:1995iw}. 
     This profile can be well approximated by a power-law distribution, expressed as $\rho(r)=\rho(r_0)(r_0/r)^{\gamma_c}$, where $r$ is the distance from the GC.
     Here, we utilize the local DM density in the solar system, $\rho(r_\odot)=0.3\,{\rm GeV/cm^3}$\,\cite{Bovy:2012tw}, at a distance $r_\odot=8.46\,{\rm kpc}$ \cite{Do:2013upa} from the GC, to define the reference density, $\rho(r_0)=\rho(r_\odot)\left({r_\odot}/{r_0}\right)^{\gamma_c}$. The value of the cusp parameter $\gamma_c$ is typically determined through numerical simulations. Simulations focusing on DM distribution reveal a range for $\gamma_c$ which is $0.9\lesssim\gamma_c\lesssim1.2$~\cite{Diemand:2008in, Navarro:2008kc}.
     $\gamma_c$ can be even larger in case of adiabatic contraction of galactic halos due to baryonic infall \cite{Blumenthal:1985qy, Gnedin:2004cx, Gustafsson:2006gr}. 
     
     \item The spike in DM density appears
     at or below $r_b$ ($r\lesssim r_b$) where $r_b\simeq 0.2$\,pc, $r_{\bf h}=0.3$\,pc~\cite{Merritt:2003qk, Merritt:2003qc}. Here we have considered $M=4\times10^6M_\odot$ for Sgr\,A$^*$ and fixed\footnote{DM infall during the formation of spike alters the velocity dispersion making it $r$-dependent (see, e.g., \cite{Fields:2014pia}). However, in most cases, our results are dominated by $s$-wave annihilation channels 
     %(especially except BMP 1, 4),
     thus will only show mild dependence. Thus, for simplicity, the velocity dispersion can be taken as constant.} $v_0=105$\,km\,s$^{-1}$ to evaluate $r_{\bf{h}}$.  
     The spike is well described by the power law $\rho_{\rm sp}(r)=\rho(r_b)(r_b/r)^{\gamma_{\rm sp}}$ where $\rho(r_b)=\rho(r_0)(r_0/r_b)^{\gamma_c}$. The value of spike parameter $\gamma_{\rm sp}$ depends on the formation history of SMBH and spike~\cite{Gondolo:1999ef, Nakano:1999ta, Ullio:2001fb, Merritt:2003qk, Gnedin:2003rj, Merritt:2006mt, Bertone:2005hw,Balaji:2023hmy}. In case of an adiabatic growth of the DM halo density around an SMBH, the $\gamma_{\rm sp}$ is usually large and takes value $\gamma_{\rm sp}^{\rm ad}=\left(\frac{9-2\gamma_c}{4-\gamma_c}\right)$\,\cite{Gondolo:1999ef, Ullio:2001fb}.
     A shallower spike implies a relatively smaller $\gamma_{\rm sp}$, which is also possible in the presence of gravitational heating by the nuclear star cluster surrounding the SMBH\,\cite{Gnedin:2003rj, Bertone:2005hw, Shapiro:2022prq, Merritt:2006mt}. 
     
     \item As we approach the SMBH from the tail of the spike, the DM density $\rho_{\rm sp}$ increases until it reaches a critical point
     at $r=r_{\rm in}$. At this critical point $\rho_{\rm sp}\sim \rho_{\rm ann}$ with $\rho_{\rm ann}={m_{\tilde\chi_1^0}}/{\langle\sigma v\rangle\tau}$ being the ``annihilation plateau" density. This forms due to a strong DM annihilation rate, triggered by very high DM density at the innermost region \cite{Gondolo:1999ef}. Here, $\tau\sim 10^{10}$\,yrs indicates the age of the SMBH. This critical density is attained at a distance $r_{\rm in}=r_b\cdot(\rho_b/\rho_{\rm ann})^{\sfrac{1}{\gamma_{\rm sp}}}$ for 
     $\rho_b =\rho(r_b)$, and for $r<r_{\rm in}$, the free rise of DM density spike towards the centre of SMBH is essentially restricted. In particular, 
     the shape of DM density gets flattened for $r<r_{\rm in}$ as characterized by a mild power law, $\rho_{\rm in}(r)=\rho_{\rm ann}\cdot(r_{\rm in}/r)^{\gamma_{\rm in}}$ with $\gamma_{\rm in}=0.5$~\cite{Vasiliev:2007vh, Shapiro:2016ypb}. 
     %\textbf{This is related to the fact close to SMBH, i.e., $r < r_{\rm in}$ the DM distribution forms a weak cusp for s-wave annihilation.}

%%%%%%%%%%%%%%%%%%%%%%%%%%%%%%%%%%%%%%%%%%%
\begin{figure}[ht]
    \includegraphics[width=0.544\textwidth]{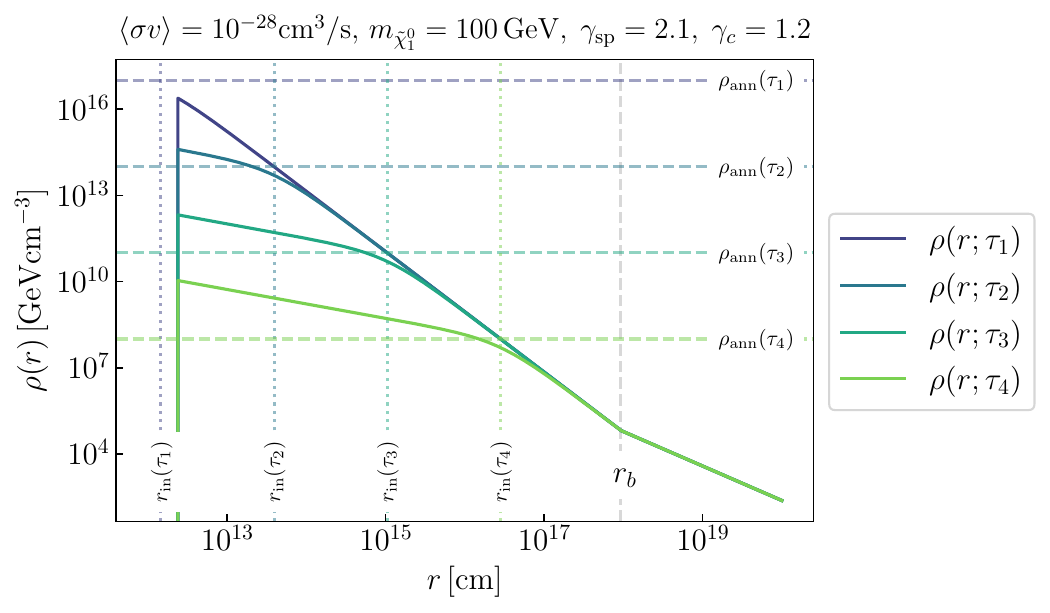}
    \includegraphics[width=0.432\textwidth]{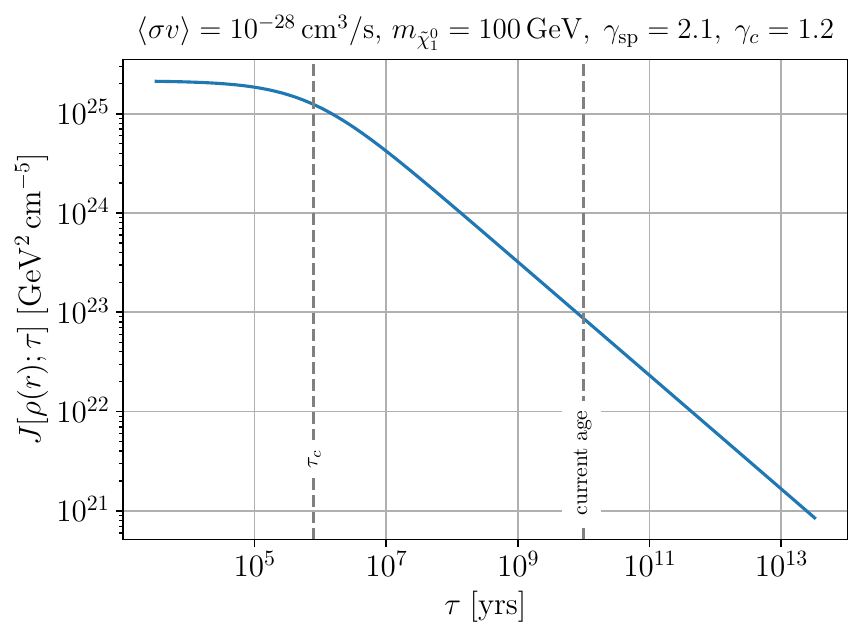}
    \caption{Left: Shape of the $\rho(r)$, considering different values for the lifetime ($\tau$) of DM spike. The values of $\tau$ are in increasing order ($\tau_1<\tau_2<\tau_3<\tau_4$). Right: $J$-factor dependence on the age of the spike. The $J$-factor remains stationary for $\tau\lesssim\tau_c$ due to little to no depletion of the spike and reduces thereafter (see the text).}
    \label{fig:rho_tau}
\end{figure}
%%%%%%%%%%%%%%%%%%%%%%%%%%%%%%%%%%%%%%%%%%%%

     \item Within distances less than $4GM$ (region inside marginally bound orbits), the gravitational influence of the SMBH is significant enough to capture all the dark matter. Consequently, little or no dark matter is present inside the sphere of radius $4GM$. 
\end{itemize}
Putting everything together, the DM density profile takes the following form: 
%%%%%%%%%%%%%%%%%%%%%%%%%%%%%%%%%%%%%%%%%%%%%%%
\begin{align}\label{eq:spike}
    \rho(r)=\begin{cases}
     0 & (r<4 G M),\\
     \dfrac{\rho_{\rm sp}(r)\rho_{\rm in}(r)}{\rho_{\rm sp}(r)+\rho_{\rm in}(r)} & (4GM\leq r\leq r_b),\\[13pt]
     \rho_b\left(\dfrac{r_b}{r}\right)^{\gamma_c} & (r_b< r\leq r_\odot),
    \end{cases}
\end{align}
%%%%%%%%%%%%%%%%%%%%%%%%%%%%%%%%%%%%%%%%%%%%%%%%%
where $\rho_{\rm sp}(r)=\rho(r_b)(r_b/r)^{\gamma_{\rm sp}}$, $\rho_{\rm in}(r)=\rho_{\rm ann}\cdot(r_{\rm in}/r)^{\gamma_{\rm in}}$ and $\rho_b=\rho(r_b)$ as defined earlier.
The spike profile is defined in such a way that a smooth transition in the DM density occurs between $r<r_{\rm in}$ and $r_{\rm in}\leq r<r_b$ regions \cite{Fields:2014pia, Shelton:2015aqa, Chiang:2019zjj}, which, however, is not always a strict requirement for the proposition of Eq.\eqref{eq:spike} \cite{Chan:2022gqd, Balaji:2023hmy}.

  In Fig.\,\ref{fig:rho_tau}\,(left), we depict the shape of $\rho(r)$ (see Eq.(\ref{eq:spike})) considering different values for the age of the DM spike. For representative purpose, we set $\langle\sigma v\rangle=10^{-28}$\,cm$^3$/s, $m_{\tilde\chi_1^0}=100$\,GeV, $\gamma_c=1.2$ and $\gamma_{\rm sp}=2.1$. Earlier we mentioned that $\rho_{\rm ann}\propto {1}/{\tau}$ implying $r_{\rm in}\sim \tau^{\sfrac{1}{\gamma_{\rm sp}}}$ and $\rho_{\rm in}(r)\sim\tau^{\sfrac{\gamma_{\rm in}}{\gamma_{\rm sp}}-1}$. These distinct features are highlighted in Fig.\,\ref{fig:rho_tau}, where we find a larger $r_{\rm in}$, corresponding to a larger $\tau$, thus leading to the early beginning of flattening of $\rho(r)$. At earlier stages (small $\tau$) following the SMBH formation $\rho_{\rm ann}$ was large, beyond the reach of $\rho_{\rm sp}(r)$ in the region $4GM<r<r_{b}$. As a result, $r_{\rm in}$ remains smaller than $4GM$ and the saturation region is not significantly manifested. It is only when $\rho_{\rm ann}$ becomes comparable to $\rho_{\rm sp}^{\rm max}\equiv\rho_{\rm sp}(r=4GM)$, and therefore, $r_{\rm in}\sim4GM$, the saturation region starts to become significant. The time scale at which this occurs is given by $\tau_c\equiv\frac{m_{\tilde\chi_1^0}}{\langle\sigma v\rangle\rho_{\rm sp}^{\rm max}}$ (see Fig.\,\ref{fig:rho_tau} where, $\tau_c\approx8\times10^5$\,yrs). For $\tau>\tau_c$, $\rho_{\rm ann}$ further reduces gradually, increasing $r_{\rm in}$. Thus, the depletion strengthens over time, enlarging the saturation region in the profile. As a consequence of this change in the profile, the $J$-factor remains stationary for $\tau\lesssim\tau_c$ (Fig.\,\ref{fig:rho_tau}\,{(right)}).
  Additionally, a higher value of the neutralino mass $m_{\tilde\chi_1^0}$ also leads to an increased $\rho_{\rm ann}$, which reduces $r_{\rm in}$, resulting in a larger $J$-factor and consequently enhances the photon flux.
  
\begin{figure}[ht]
    \centering
   \includegraphics[width=0.45\linewidth]{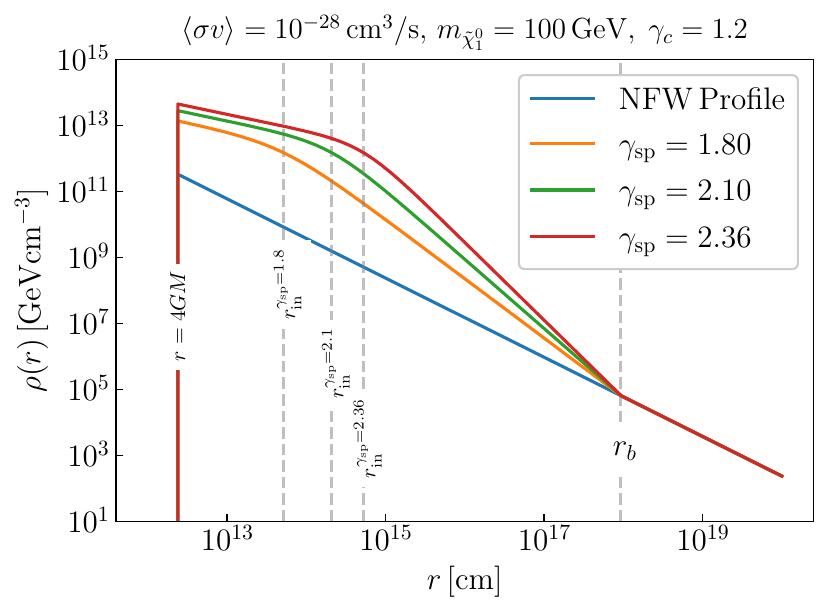}
   \includegraphics[width=0.45\linewidth]{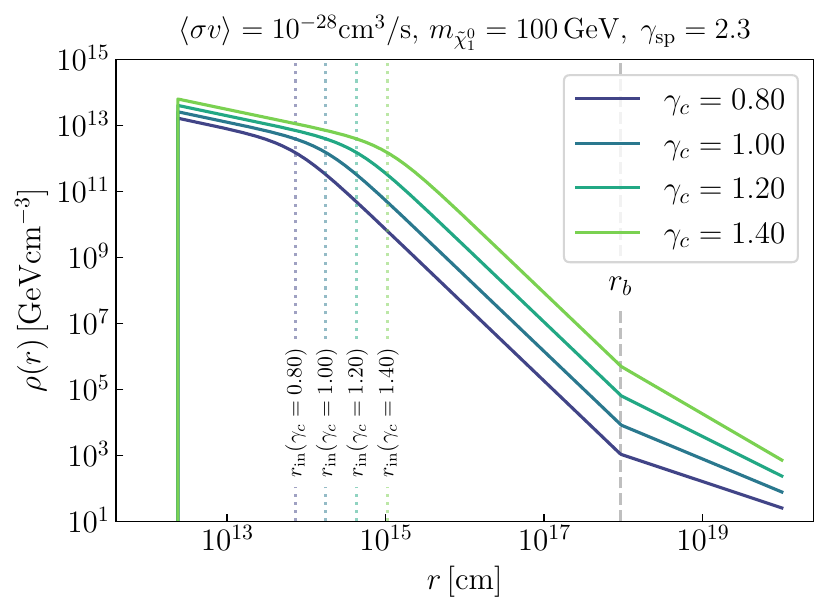}
   \caption{$\rho(r)$-$r$ behavior for varying $\gamma_{\rm sp}$ (left) and $\gamma_c$ (right).
   Here, $\rho_{\rm ann}$ is evaluated considering $\tau\sim10^{10}$\,yrs, the typical age of the DM spike associated with the SMBH.}
    \label{fig:figSpike}
\end{figure}

  Next, in Fig.\,\ref{fig:figSpike}, we discuss the sensitivity of $\gamma_{\rm sp}$ (left) and $\gamma_{c}$ (right) to $\rho(r)$. Once again, we consider the same representative choices of $m_{\tilde\chi_1^0}$ and $\langle\sigma v\rangle$ for the purpose.
  Additionally, we fix $\gamma_c$ in the left plot
  and $\gamma_{\rm sp}$ in the right plot and consider $\tau=10^{10}$\,{\rm yrs}. 
  We find that a larger $\gamma_{\rm sp}$ results in a steeper spike, specifically, in the intermediate region $r_{\rm in}<r < r_b$. Similarly, a larger $\gamma_c$ leads to an enhancement of $\rho(r)$ in the whole region of interest ($4GM\leq r\leq r_\odot$). Since in this study, we are aiming for a possible amplification in the $\gamma$-ray flux, resulting from DM annihilation, we set $\gamma_c$ to 1.2 (maximum allowed value from DM only simulations\, \cite{Diemand:2008in, Navarro:2008kc}) throughout the analysis and treat $\gamma_{\rm sp}$ as a free parameter. 
 
%%%%%%%%%%%%%%%%%%%%%%%%%%%%%%%%%%
%%%%%%%%%%%%%%%%%%%%%%%%%%%%%%%%%%%

In the present work, the resulting $\gamma$-ray  flux from DM annihilation dominantly originates from a narrow region containing the density spike~\cite{Chiang:2019zjj}. 
Here, we will treat the DM spike as a $\gamma$-ray point source. The contribution from the spike to the $J_{\rm halo}$ is given as \cite{Chiang:2019zjj},
\begin{align}
    J[\rho(r)]&=\frac{1}{r_\odot^2}\int_{4GM}^{r_b} dr\,\rho(r)^2 r^2 .\label{eq:Jsp}
\end{align}
One finds that, for the relevant region of interest, the $J$-factor from the spike, appearing in Eq.(\ref{eq:Jsp}), is 3 to 6 orders larger than the entire $J_{\rm halo}$ in the absence of the spike, depending upon $m_{\tilde\chi_1^0}, \langle\sigma v\rangle$ and $\gamma_{\rm sp}$. Thus, the differential photon flux in the presence of the DM spike in Eq.(\ref{eq:ordFlux}) can be approximated as,
%\cite{Chiang:2019zjj},

\begin{align}
    \dfrac{d\Phi}{dE_\gamma}&\simeq\dfrac{\langle\sigma v \rangle}{2m_{\tilde\chi_1^0}^2}\dfrac{dN}{dE_\gamma}\times J\left[\rho(r)\right].\label{eq:dPhidE}
\end{align}

\begin{figure}[ht]
    \centering
    \includegraphics[width=0.5\linewidth]{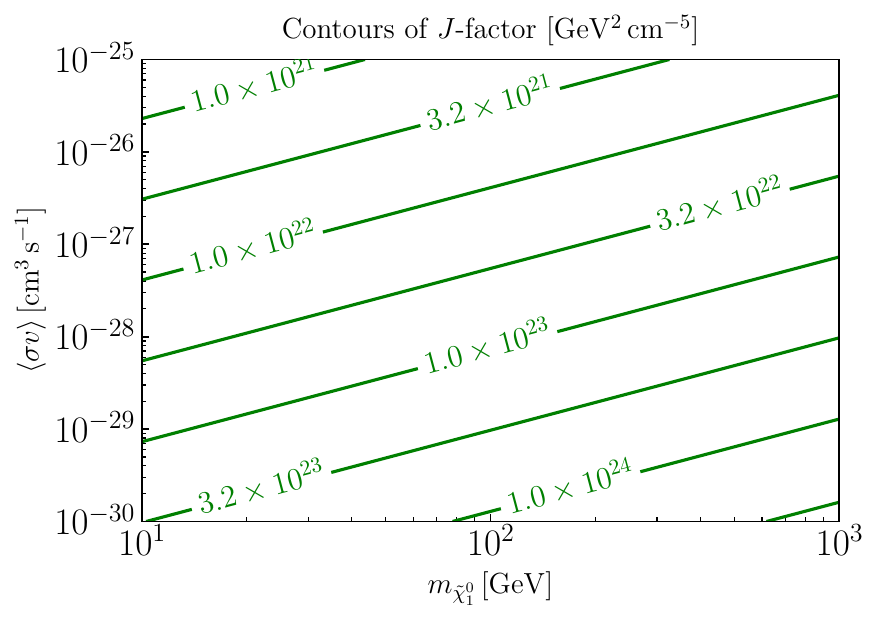}
    \caption{Contours of $J$-factor values in $m_{\tilde\chi_1^0}\,-\,\langle\sigma v\rangle$ plane for the DM density profile of Eq.(\ref{eq:spike}) with 
    $\gamma_{\rm sp}=2.1$ and $\gamma_c=1.2$
    .} 
    \label{fig:JFact}
\end{figure}

Unlike the case of a conventional DM ({\it e.g. }NFW) halo profile where the $J$-factor can be segregated from the particle physics inputs, interestingly, for the DM density profile of Eq.(\ref{eq:spike}), the $J$-factor turns sensitive to the DM annihilation cross-section and the DM mass. 
This is demonstrated in Fig.\,\ref{fig:JFact} where the contours of $J$-factor are plotted in $m_{\tilde{\chi}_1^0}-\langle\sigma v\rangle$ plane. A rise in the $J$ values can be observed 
with a low $\langle\sigma v \rangle$ and for a high DM mass.

%%%%%%%%%%%%%%%%%%%%%%%%%%%%%%%%%%%%%

\begin{figure}
     \centering
     \includegraphics[width=0.5\linewidth]{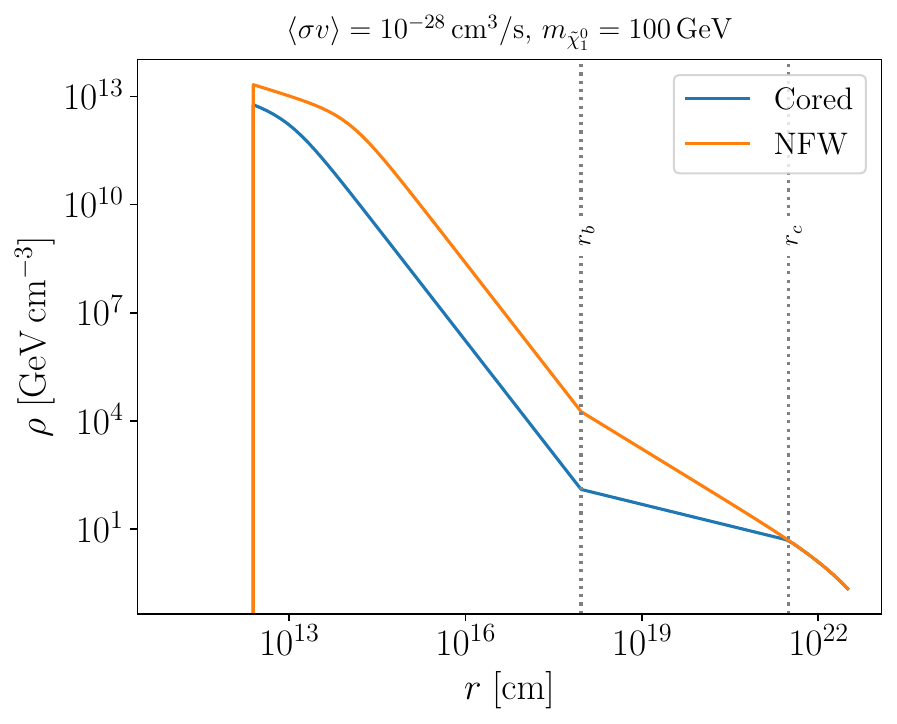}
     \caption{Spike profiles for cored ($\gamma_c=0.4$, $r_c=1$\,kpc, $\gamma_{\rm sp}=2.1$) and NFW ($\gamma_c=1.2$, $\gamma_{\rm sp}=2.1$).}
     \label{fig:core_NFW}
 \end{figure}

 Before we conclude our general discussion on photon flux, we note that the spike heavily depends on whether the halo profile is cored or cuspy, characterized by the radial extension ($r_b$) of the DM spike and the core radius $r_c$.
 \begin{align}
 \rho^{\rm sp}_{\rm NFW}(r)=
 \begin{cases}
      \rho_{\rm NFW}(r) & (r>r_b),\\[8pt]
    \dfrac{\rho_{\rm sp}(r)\rho_{\rm in}(r)}{\rho_{\rm sp}(r)+\rho_{\rm in}(r)} & (4GM\leq r\leq r_b),\\[8pt]
     0 & (r<4 G M),
     \end{cases}
     \;\;\;\;\vline& \;\;\rho_{\rm core}^{\rm sp}(r)&=
     \begin{cases}
         \rho_{\rm NFW}(r)&r\ge r_c,\\
         \rho_{\rm NFW}(r_c)\left(\dfrac{r_c}{r}\right)^{\gamma_c} & r_b\le r <r_c,\\[8pt]
        \dfrac{\rho_{\rm sp}(r)\rho_{\rm in}(r)}{\rho_{\rm sp}(r)+\rho_{\rm in}(r)} & (4GM\leq r\leq r_b)
        ,\\[8pt]
         0 & (r<4 G M),
     \end{cases} 
 \end{align}
For our galaxy, simulations
allow for core radii of the order of a 
kpc and $0\leq \gamma_c <1$~\cite{Balaji:2023hmy}.
Similarly, the Navarro-Frenk-White (NFW) profile is given as~\cite{Navarro:1995iw}:
 \begin{align}
     \rho_{\rm NFW}(r)=\dfrac{\rho_0}{\frac{r}{R_s}\left(1+\frac{r}{R_s}\right)^2}&&\rho_0=\rho(R_\odot)\left[\frac{R_\odot}{R_s}\left(1+\frac{R_\odot}{R_s}\right)^2\right]
 \end{align}
 where $R_s=18.6$\,kpc is the scale radius, and $\rho_0$ is the density parameter obtained using local density at the sun ($\rho(R_\odot)=0.3$\,GeV/cm$^3$) as reference. The distinction between the spikes of the two profiles is clear from Fig.\,\ref{fig:core_NFW}. Here $\gamma_{\rm sp}=2.1$ and the other parameters are taken from Ref.~\cite{Balaji:2023hmy}. 
 %Considering the NFW profile for the DM halo,
 
 As discussed in ref.\,\cite{Fields:2014pia}, we note that spike profiles may become shallow inside cored halos. This may arise from mergers, strong supernova feedbacks, etc. As mentioned earlier, the nature of the spike may also depend on the formation history of the Milky Way for its SMBH, such as an influence of a compact object or DM self-interactions. Additionally, we assumed photon signal emissions to arise entirely from DM annihilation, disregarding any other sources like active galactic nuclei (AGN). Refs.\,\cite{Ullio:2001fb,Balaji:2023hmy} also discuss various other uncertainties related to the spike based on different considerations.
 
%%%%%%%%%%%%%%%%%%%%%%%%%%%%%%%%%
\section{Precision Observables and Constraints from colliders and DM Searches}\label{sec:3}

We outline here the necessary steps followed to evaluate the differential $\gamma$-ray
flux ($d\Phi/dE_\gamma$) and the required spike for the $\tilde \chi_1^0$ DM considering the density profile as given in Eq.\eqref{eq:spike}.
We use the publicly available code \textsf{SARAH 4.15.2} \cite{Staub:2013tta, Staub:2015kfa} for generating the MSSM model files for both \textsf{SPheno} \cite{Porod:2003um, Porod:2011nf} and \textsf{MicrOMEGAs} \cite{Belanger:2001fz, Belanger:2013oya, Belanger:2006is, Belanger:2008sj, Alguero:2023zol}. Next, using \textsf{SPheno 4.0.5} \cite{Porod:2003um, Porod:2011nf}, we calculate the particle spectrum that is fed into \textsf{MicrOMEGAs 6.0}\,\cite{Alguero:2023zol} which further computes the DM relic abundance and SI-DD cross-section. The quantities, ${dN}/{dE_\gamma}$ and $\langle\sigma v\rangle$, relevant for computing $\gamma$-ray flux in Eq.(\ref{eq:dPhidE}), are also subsequently obtained from \textsf{MicrOMEGAs 6.0} \cite{Alguero:2023zol}. The $J$-factor is calculated using Eq.(\ref{eq:Jsp}). Finally, we follow the steps discussed in Sec.\,\ref{Sec:2} to calculate the photon flux for the MSSM parameter space.

The validity of each point is tested through the relevant observables related to precision and collider phenomenology. If the SUSY states are sufficiently light, they would impact various precision observables. Here, we enlist the constraints from different precision experiments relevant to our analysis. Unless otherwise stated, all these observables have been calculated using \textsf{SPheno 4.0.5} \cite{Porod:2003um, Porod:2011nf}.

$\bullet$ The light, neutral, CP-even Higgs boson resembles the SM Higgs with a theoretical uncertainty of 3\,GeV around the central value of 125\,GeV \cite{CMS:2012qbp, ATLAS:2012yve}. Thus the range for $m_h$ can be read as,
\begin{equation}\label{eq:Hmass}
122\,{\rm GeV}<m_h<128 \,\rm GeV.
\end{equation}  

Additionally, it is crucial to test the branching ratios of the Higgs scalar predicted by the model against the limits from the LHC. We use \textsf{HiggsTools}~\cite{Bahl:2022igd, Bechtle:2020pkv, Bechtle:2008jh, Bechtle:2020uwn, Bechtle:2013xfa} to check for such constraints. The package includes \textsf{HiggsBounds\,-\,5.10.2} \cite{Bechtle:2020pkv, Bechtle:2008jh} and \textsf{HiggsSignals\,-\,2.6.2} \cite{Bechtle:2020uwn, Bechtle:2013xfa}.

$\bullet$ 
As noted already, the recent experiment, E989 at the Fermilab, provides new results from the Run-2 and Run-3 data analyses, providing the new deviation of the anomalous magnetic moment of the muon,
\cite{Muong-2:2021ojo,fermi2023},
\begin{align}
    \delta a_\mu=a_\mu^{\rm exp}-a_\mu^{\rm SM}=(249\pm 48)\times 10^{-11},
    \label{eq:delamu}
\end{align} 
compared to that of the SM value. If the entire discrepancy stems from SUSY, then for a valid parameter space point $a_\mu^{\rm SUSY}$ needs to reconcile with $\delta a_\mu$.
In the above $a_\mu^{\rm SM}=(116591810 \pm 43)\times 10^{-11}$ \cite{Aoyama:2020ynm} has been considered (For the SM calculations, see also \cite{Davier:2017zfy, Keshavarzi:2018mgv, Colangelo:2018mtw, Davier:2019can, Keshavarzi:2019abf, Kurz:2014wya, Hoferichter:2019mqg, Melnikov:2003xd, Masjuan:2017tvw, Colangelo:2017fiz, Hoferichter:2018kwz, Gerardin:2019vio, Bijnens:2019ghy, Colangelo:2019uex, Colangelo:2014qya, Blum:2019ugy, Aoyama:2012wk, atoms7010028, Czarnecki:2002nt, Gnendiger:2013pva}). 
The combined experimental result for the value of $a^{\rm exp}_\mu$ deviates from the SM predicted value by $5.1\sigma$. The Standard Model's predicted value, $a_\mu^{\rm SM}$, is subject to theoretical uncertainty primarily due to the leading-order contribution from Hadronic Vacuum Polarization (HVP). In this context, $a_\mu^{\rm SM}$ is determined using HVP contributions derived from $e^+e^-\rightarrow$ hadrons data through dispersive techniques~\cite{Aoyama:2020ynm}, $a_\mu^{\rm HVP} \big\lvert_{e^+e^-}=(6931\pm40)\times 10^{-11}$. However, utilizing HVP contributions from lattice-QCD results by the BMW collaboration,
$a_{\mu}^{\rm HVP}\big\lvert_{\rm BMW}= (7075\pm55)\times 10^{-11}$~\cite{Borsanyi:2020mff} reduces the discrepancy in $a_\mu$ to $1.6\sigma$.
Referring the variation as $\delta a_\mu^{\rm BMW}$,
it reads 
\begin{align}
    \delta a_\mu^{\rm BMW}=(107\pm69)\times10^{-11}~.
\label{eq:bmw}
\end{align}

Other lattice groups~\cite{Ce:2022kxy, ExtendedTwistedMass:2022jpw, FermilabLatticeHPQCD:2023jof} show some agreement with the BMW result, highlighting a notable disagreement among the various estimates of HVP contributions. This discrepancy underlines unresolved issues inherent to each methodology.

%ref. \cite{Wittig:2023pcl} discusses this tension without considering the recent CMD3 result. 
The use of Euclidean time windows to reconcile discrepancies between lattice QCD and the $e^+e^-\to$ hadrons cross-section data regarding the HVP contribution to $(g-2)\mu$ is explored in~\cite{Colangelo:2022vok}. Reference~\cite{Wittig:2023pcl} also emphasizes the tension between the lattice QCD method and traditional data-driven approaches, noting that the recent CMD-3 result was not included in the latter. 
More recent work, Ref.~\cite{CMD-3:2023rfe}, calculates $a^{\rm SM}_\mu$ using a data-driven approach for HVP contributions, incorporating CMD3 data which leads to $0.9\sigma$ for $\delta a_\mu$. We subsequently denote the variation as $\delta a_\mu^{\rm CMD3}$.
\begin{align}
    \delta a_\mu^{\rm CMD3}=(49\pm55)\times10^{-11}~.
\label{eq:cmd3}
\end{align}
The recent data appear to be inconsistent with previous findings~\cite{CMD-3:2023alj, CMD-2:2003gqi, KLOE-2:2017fda, BESIII:2015equ}. This reveals significant differences in the Hadronic Vacuum Polarization (HVP) results across these investigations, indicating a need for further research to resolve these discrepancies~\cite{Colangelo:2022jxc}. 

In MSSM, the leading contribution to $a^{\rm SUSY}_\mu$ comes from the neutralino-smuon and chargino-sneutrino loops at one loop level. 
Under the assumption that all sparticles have a uniform mass $M_{\rm SUSY}$, an approximated form of $a^{\rm SUSY}_\mu$ at one-loop is given by\,\cite{Moroi:1995yh, Martin:2001st},
\begin{align}\label{eq:g21loop} a^{\rm SUSY}_\mu=\frac{1}{192\pi^2}\frac{m_\mu^2 \tan\beta}{M_{\rm SUSY}^2}(5 g_2^2+g_1^2),
\end{align}
where $g_2$ and $g_1$ are the gauge coupling constants corresponding to $SU(2)_L$ and $U(1)_Y$ respectively.
The Eq.(\ref{eq:g21loop}) gives the impression that $a^{\rm SUSY}_\mu$
contribution in MSSM can be enhanced with a relatively larger $\tan\beta$ and lighter SUSY particles, specifically having the weak interaction.

We note that
$a_\mu^{\rm SUSY}$ may also receive 
contributions from the two-loop (mainly Barr-Zee type) diagrams involving fermion/sfermion in the loop~\cite{Fargnoli:2013zda, Fargnoli:2013zia}.
Hence, we use the publicly available package \textsf{GM2Calc}\,\cite{Athron:2015rva, Athron:2021evk, vonWeitershausen:2010zr, Bach:2015doa, Fargnoli:2013zia, Fargnoli:2013zda}, which computes the $a_\mu^{\rm SUSY}$ value up to two loops. We generate the input file using \textsf{SPheno 4.0.5} and pass the tree-level masses to \textsf{GM2Calc} \cite{Athron:2015rva, Athron:2021evk, vonWeitershausen:2010zr, Bach:2015doa, Fargnoli:2013zia, Fargnoli:2013zda} for obtaining $a_\mu^{\rm SUSY}$. In delineating our
results in Sec.\eqref{sec:4} we use Eq.\eqref{eq:delamu}, Eq.\eqref{eq:bmw} and
Eq.\eqref{eq:cmd3}, considering the different estimations of $a_\mu^{\rm HVP}$. Since BMW
or CMD3 collaborations hint towards a weaker difference between SM prediction and experimental observations of $a_\mu$,
$a_\mu^{\rm SUSY}$ provides only a mild restriction on the SUSY parameter space. It mainly allows heavier Higgsino-like states and
smuons, while the lighter EW states now become constrained. 

$\bullet$ Assuming the $\tilde{\chi}_1^0$ to be the only source for DM, 
the acceptable value of the relic abundance data~\cite{WMAP:2012nax, Planck:2018vyg} at $1\sigma$ confidence level is read as %
 \begin{equation}
   \Omega_{\rm DM}h^2=0.1198\pm 0.0012.
   \label{eq:relic}
 \end{equation}
\noindent

We consider a $2\sigma$ level of allowance of $ \Omega_{\rm DM}h^2$ in Eq.\eqref{eq:relic}, thus accept the parameter points  
of both $\tilde B_{\tilde H}$ and $\tilde B_{\tilde W\tilde H}$ 
types that satisfy $0.1174< \Omega_{\rm DM}h^2
< 0.1222$.\footnote{We assume a conventional radiation-dominated universe and adopt the standard thermal freeze-out mechanism for the DM relic abundance.} 

The spin-independent scattering of neutralino DM\, and the light quark and gluons can be read from the following effective Lagrangian \cite{Drees:1993bu, Nath:1994ci, Hisano:2004pv, Hisano:2017jmz, PhysRevD.59.055009}
\begin{align}
    \mathcal{L}_{\rm SI}^{\rm eff}=f_q \tilde{\chi}_1^0 \tilde{\chi}_1^0 \overline{q}q+f_G{\tilde{\chi}}^0_1\tilde{\chi}^0_1G^a_{\mu\nu}G^{a\,\mu\nu} + \dots
\label{eff_lag}
\end{align}
where, the effective couplings $f_q$ and $f_G$ include the effects of the SI scattering, mediated by the light quarks and gluons, respectively. The first term in the R.H.S. receives contributions mainly from the exchange of CP-even Higgses via $t$-channel. The squark exchange contribution is subdominant and can be neglected in this analysis. 
The twist-2 operators (traceless part of the energy-momentum tensor)
for the quarks and gluons may also be present in Eq.\eqref{eff_lag} as indicated by the ellipsis.
However, we do not include any new twist-2 contributions except those already present in \textsf{MicrOMEGAs} \cite{Belanger:2008sj}.

The SI scattering cross-section of the neutralino with target nuclei can be expressed as \cite{Jungman:1995df, Ellis:1987sh, Barbieri:1988zs, Ellis:2000ds, Vergados:2006sy, Oikonomou:2006mh, Ellis:2008hf, Nath:1994ci}
\begin{align}
\sigma_{\rm SI}=\frac{4}{\pi}\left(\frac{m_{\tilde{\chi}^0_1}M_A}{m_{\tilde{\chi}^0_1}+M_A}\right)^2\Bigg[\{Zf_p+(A-Z)f_n\}^2
\Bigg],
\label{eq:sigma_nucl}
\end{align} 
where 
$Z$ and $A$ represent its atomic and mass numbers, respectively. The form factor $f_N$ ($N=p,n$) in Eq.\eqref{eq:sigma_nucl} 
includes spin-independent coupling of the neutralino with nucleon (of mass $m_N$) \cite{Drees:1993bu, Nath:1994ci, Hisano:2004pv, Hisano:2017jmz, PhysRevD.59.055009}.
In the numerical representation, we have used the recent results of LUX-ZEPLIN (LZ) collaboration\,\cite{LZ:2022lsv} on the SI-DD cross-section, which gives a stronger upper limit than the previous XENON1T\,\cite{XENON:2018voc}, PandaX-4T\,\cite{PandaX-4T:2021bab}, LUX\,\cite{LUX:2016ggv} experiments to constrain our model parameter space.

 $\bullet$ The presence of a light SUSY state can significantly alter
the branching fractions (Br) of certain flavour-violating processes, e.g. $B\to X_s\gamma$, $B_s\to \mu^+\mu^-$, $B^+\to \tau^+\nu_\tau$ and $B\to X_s\nu\,\overline{\nu}$  compared to the SM
predictions.  We consider the following bounds for our analysis,
\begin{align}
& {\rm Br}(B\to X_s\gamma)\,=\,(3.49\pm 0.38)\times 10^{-4},\\
& {\rm Br}(B_s^0\to \mu^+\mu^-)\,=\,(3.45\pm 0.58)\times 10^{-9},  \\
& {\rm Br}(B^+\to \tau^+\nu_\tau)\,=(1.09\pm 0.4)\times 10^{-4}\,, 
\end{align}
at 95$\%$ C.L.\,\cite{ParticleDataGroup:2022pth,hflavichep2022}. 
Similarly, we used 
\begin{align}
& {\rm Br}(B\to X_s\nu\,\overline{\nu})\,<\, 6.4\times 10^{-4}, 
\end{align}
in this analysis \cite{Yamada:2007me}.

$\bullet$ Searches for the lighter charginos, lighter neutralinos, and sleptons at a center-of-mass energy of $\sqrt{s}=13$\, TeV through various final states like $2l +jets+\met$, $3l+jets+\met$, etc. (here, $l$ stands for leptons of flavours $e$ or $\mu$) have been conducted by both the ATLAS \cite{ATLAS:2018ojr, ATLAS:2018eui, ATLAS:2019lff, ATLAS:2019lng, ATLAS:2020pgy, ATLAS:2019wgx, ATLAS:2021moa, ATLAS:2024lda, ATLAS:2024qxh, ATLAS:2024fub, ATLAS:2022zwa} and the CMS \cite{CMS:2018kag, CMS:2018szt, CMS:2018eqb, CMS:2020bfa, CMS:2021edw, CMS:2017moi} collaborations of the LHC in the context of simplified models.  
Although in this work only compressed-type scenarios \cite{ATLAS:2019lng} are relevant, for completeness, we also mention the mass bounds at LHC Run-II for non-compressed models.
Ref. \cite{Adam:2021rrw} includes a more recent review on electroweakinos at the LHC.

In Ref. \cite{ATLAS:2019lng}, ATLAS collaboration discussed the searches for lighter electroweakinos and sleptons in compressed-type scenarios where the mass gaps between electoweakinos/sleptons and the LSP are very small. The bounds are drastically relaxed in such cases as the final state products are very soft, making their experimental detection more challenging.
A brief discussion on LHC searches for such sparticles through various lepton multiplicity channels is given below :

\begin{itemize}

    \item The final state comprising $2l+jets+\met$ \cite{ATLAS:2019lff} arising from the production process $pp\rightarrow \tilde\chi_1^\pm \tilde\chi_1^\mp$ at LHC Run-II for an integrated luminosity of 139 fb$^{-1}$ put bounds on $m_{\tilde\chi_1^\pm }$ around 420 GeV for negligible LSP mass in which $\tilde\chi_1^\pm$ decays via on-shell SM gauge bosons. The bound becomes 1 TeV when lighter chargino decays via sleptons. In both cases, the lighter charginos are Wino-like. For pair production of sleptons, the bound is about 700 GeV for negligible LSP mass.
    For details, see ref. \cite{ATLAS:2019lff}.
    
    \item  At LHC Run-II with an integrated luminosity of 139\,fb$^{-1}$, the ATLAS collaboration \cite{ATLAS:2019lng} has obtained bound on the mass of $\tilde \chi_1^{\pm}/\tilde \chi_2^0$ around 640\,GeV for negligible LSP mass through the associated production process, $pp\rightarrow \tilde\chi_2^0 \tilde \chi_1^\pm $, which results in the final state containing $3l+jets+\met$. Here, $\tilde \chi_1^\pm/\tilde \chi_2^0$ decays via on-shell SM gauge bosons ($W^\pm/Z$). For off-shell SM gauge bosons ($W^\pm/Z$), the corresponding bound is 300 GeV. If $\tilde \chi_2^0$ decays into the Higgs boson and the LSP, the corresponding bound on $\tilde\chi_2^0 /\tilde \chi_1^\pm$ reduces to 190 GeV. In all these cases,  $\tilde\chi_2^0/\tilde \chi_1^\pm $ are assumed to be Wino-like. However, for Higgsino-type scenarios the corresponding bound on $m_{\tilde\chi_2^0 }$ is 210 GeV for almost degenerate $\tilde\chi_1^0$,$\tilde\chi_2^0$ and $\tilde \chi_1^\pm$. For details, see the refs. \cite{ATLAS:2021moa}. 
    
    \item For compressed-type scenarios (pertinent for this work) the ATLAS collaboration \cite{ATLAS:2019lng} puts bounds on masses of lighter electoweakinos and sleptons through the productions, $pp\rightarrow \tilde\chi_2^0 \tilde \chi_1^\pm j$ and $pp\rightarrow {\tilde l}\, {\tilde l}^* j$ respectively at LHC Run-II where, $j$ stands for $jet$. These processes finally lead to a signal, $2l+jet+\met$, with the opposite sign, the same flavour leptons having soft 
    transverse momentum
    ($p_T$) with a boosted jet. The production of lighter electroweakinos ($\tilde \chi_1^\pm, \tilde\chi_{2,3}^0$) in association with two boosted forward jets via vector boson fusion processes has a low cross-section, but due to the presence of boosted jets, it has improved signal-to-background ratio, signal isolation and identification power. These improvements lower the systematic uncertainties while enhancing the signal sensitivity.
    
    \item For Higgsino-type models, the bound on lighter electroweakinos are around 193 GeV for a mass splitting of 9.3 GeV between $\tilde\chi_2^0/\tilde \chi_1^\pm $ and the LSP. For Wino-type scenarios, the corresponding mass bound is 240 GeV for the mass splitting of 7 GeV of the same set of sparticles. In compressed scenarios, the bound on masses of sleptons is around 251 GeV for 10 GeV mass splitting between sleptons and the LSP. In this analysis, events consistent with the production of electroweakinos through vector-boson fusion processes are used to constrain Wino and Higgsino models via $q \bar q $ fusion production processes. 
    Here, additional hadronic initial-state radiation(ISR) jets are required at the production level to enhance the sensitivity of the signal.
    For details, see the ref. \cite{ATLAS:2019lng}.
    \item Electroweakino searches have also been performed at the LHC Run-II experiment in the $1l+2\,b\,\text{-}jets+\met$ channel \cite{ ATLAS:2020pgy} for the production process $pp\rightarrow \tilde\chi_2^0 \tilde \chi_1^\pm $. In this analysis, they assumed the BR of $\tilde \chi_2^0\rightarrow \tilde\chi_1^0 h$ to be 100$\%$ where $h$ decays into a pair of $b$-quarks, finally leading to one lepton signal. Here, a Wino-type scenario is considered, and the mass bound obtained is around 740 GeV for negligible LSP mass.
    \item
    The refs. \cite{ATLAS:2022hbt, ATLAS:2019lff, ATLAS:2019lng, ATLAS:2018ojr, CMS:2018eqb}     provide results on searches for sleptons and lighter electroweakinos which result in final states comprising of charged leptons ($e^\pm, \mu^\pm$) and $\tilde \chi_1^0$. 
     With $\Delta m (\tilde l, \tilde \chi_1^0) = m_{\tilde l}-m_{\tilde \chi_1^0}$ and $\Delta m (\tilde \chi_1^\pm, \tilde \chi_1^0) = m_{\tilde \chi_1^\pm}-m_{\tilde \chi_1^0}$, Ref. \cite{ATLAS:2022hbt} investigated 
     the 'moderately compressed and  experimentally challenging mass regions $\Delta m (\tilde l, \tilde \chi_1^0)$ and 
     $\Delta m (\tilde \chi_1^\pm, \tilde \chi_1^0)$ that are close to 
     the $W$-boson mass.
     The slepton masses up to 150\,GeV was excluded for $\Delta m(\tilde l,\chi_1^0)$ to be around 50\,GeV. For lighter charginos, the mass bound obtained is 140\,GeV for $\Delta m(\tilde \chi_1^\pm ,\tilde \chi_1^0)$ to be around 100\,GeV.

  \item 
  The High-Luminosity LHC (HL-LHC) will significantly enhance our ability to probe the region of parameter space corresponding to compressed SUSY. The HL-LHC is designed to operate at 14\,TeV with a much higher integrated luminosity ($3000$\,fb$^{-1}$) compared to the ongoing LHC Run-II, offering better sensitivity to scenarios like compressed supersymmetry, where the mass gap between the sparticles is small. In compressed scenarios, the mass bounds on lighter electroweakinos and sleptons of the first two generations obtained from LHC Run-II are around (190-250)\,GeV \cite{ATLAS:2019lng} depending upon the composition of electroweakinos and the particular signatures to be analyzed. However, for third-generation sleptons, it is a little weaker because of tau-tagging efficiency. For lighter charginos/second lightest neutralinos, the bounds extend up to around 350\,GeV \cite{espgbook,ATLAS:2018jjf,CMS:2018kag,CMS:2018qsc} if HL-LHC is considered. In HL-LHC, there are challenges due to enhancement in pile-up events, which may contaminate the signal significantly. The extra boosted jet originating from initial state radiation (ISR) can be a good handle to enhance the sensitivity of the signal over the background. Moreover, the detection efficiency and reconstruction efficiency being better due to the large amount of data can also increase the significance of the signal, thereby reducing various uncertainty factors. Hence, the overall effect is to extend the allowed parameter space compared to the parameter space consistent with the LHC Run-II/III data.

In this work, the BMPs considered are found to satisfy even the HL-LHC projected mass bounds~\cite{ATLAS:2018jjf,CMS:2018kag,CMS:2018qsc} on lighter electroweakinos for Higgsino or Wino type LSP in the MSSM.
Any precise prediction of the concerned parameter space in the light of HL-LHC is beyond the scope of this work. 

\end{itemize}

The latest package \textsf{SModelS 2.3.3}~\cite{Kraml:2013mwa, Ambrogi:2017neo, Ambrogi:2018ujg, Alguero:2020grj, Alguero:2021dig} is used to test the set of parameters against these constraints.

%%%%%%%%%%%%%%%%%%%%%%%%%%%%%%%%%%
\section{Results}
\label{sec:4}
%%%%%%%%%%%%%%%%%%%%%%%%%%%%%%%%%%
As mentioned earlier, in the present work, we explore two light LSP scenarios $\tilde B_{\tilde H}$ and $\tilde B_{\tilde W\tilde H}$ 
and probe the $\gamma$-ray search prospects 
around an SMBH that can also accommodate the recent $a_\mu$ data and satisfy other relevant experimental constraints, including the SI-DD bound. We begin by presenting six distinct benchmark points (BMPs) in Table\,\ref{tab:BMPs} for illustration. The first three points pertain to the $\tilde B_{\tilde H}$ case, and the remaining three represent the $\tilde B_{\tilde W \tilde H}$ scenario. All the BMPs can accommodate the correct relic abundance (via Eq.(\ref{eq:relic})) and the desired SUSY contribution to $\delta a_\mu$ in agreement with
Eq.(\ref{eq:delamu}). The chosen BMPs are also consistent with the bound on flavour violating processes, LHC constraints, limits on SI-DD cross-section and the observables that check the mass and the properties of an SM-like Higgs scalar.
Table\,\ref{tab:BMPs} shows the values of a few such observables corresponding to each BMP.
%%%%%%%%%%%%%%%%%%%%%%%%%%%%%%%%%%%%%%%
\begin{table*}[h]
  \renewcommand*{\arraystretch}{1.4}
\begin{tabular}{|l|lll|lll|}
\hline
\multirow{2}{*}{} &
  \multicolumn{3}{c|}{\textbf{$\tilde B_{\tilde H}$ }} &
  \multicolumn{3}{c|}{\textbf{$\tilde B_{\tilde W\tilde H}$}} \\ \cline{2-7} 
 &
  \multicolumn{1}{c|}{BMP\,1} &
  \multicolumn{1}{c|}{BMP\,2} &
  \multicolumn{1}{c|}{BMP\,3} &
  \multicolumn{1}{c|}{BMP\,4} &
  \multicolumn{1}{c|}{BMP\,5} &
  \multicolumn{1}{c|}{BMP\,6} \\ \hline
\multicolumn{1}{|c|}{$M_1$\,[GeV]} &
  \multicolumn{1}{c|}{200} &
  \multicolumn{1}{c|}{300} &
  \multicolumn{1}{c|}{350} &
  \multicolumn{1}{c|}{200} &
  \multicolumn{1}{c|}{300} &
  \multicolumn{1}{c|}{600} \\
\multicolumn{1}{|c|}{$M_2$\,[GeV]} &
  \multicolumn{1}{c|}{1500} &
  \multicolumn{1}{c|}{1500} &
  \multicolumn{1}{c|}{1500} &
  \multicolumn{1}{c|}{230} &
  \multicolumn{1}{c|}{302} &
  \multicolumn{1}{c|}{582} \\
\multicolumn{1}{|c|}{$\mu$\,[GeV]} &
  \multicolumn{1}{c|}{810} &
  \multicolumn{1}{c|}{800} &
  \multicolumn{1}{c|}{800} &
  \multicolumn{1}{c|}{810} &
  \multicolumn{1}{c|}{900} &
  \multicolumn{1}{c|}{1200} \\
\multicolumn{1}{|c|}{$\tan \beta$} &
  \multicolumn{1}{c|}{16} &
  \multicolumn{1}{c|}{47} &
  \multicolumn{1}{c|}{45} &
  \multicolumn{1}{c|}{16} &
  \multicolumn{1}{c|}{25} &
  \multicolumn{1}{c|}{55} \\
\multicolumn{1}{|c|}{$m^{\rm in}_{\tilde{e}_L},m^{\rm in}_{\tilde{e}_R}$\,[GeV]} &
  \multicolumn{1}{c|}{221} &
  \multicolumn{1}{c|}{335} &
  \multicolumn{1}{c|}{379} &
  \multicolumn{1}{c|}{221} &
  \multicolumn{1}{c|}{350} &
  \multicolumn{1}{c|}{624} \\
\multicolumn{1}{|c|}{$m^{\rm in}_{\tilde{\mu}_L},m^{\rm in}_{\tilde{\mu}_R}$\,[GeV]} &
  \multicolumn{1}{c|}{225} &
  \multicolumn{1}{c|}{340} &
  \multicolumn{1}{c|}{381} &
  \multicolumn{1}{c|}{225} &
  \multicolumn{1}{c|}{357} &
  \multicolumn{1}{c|}{635} \\
\multicolumn{1}{|c|}{$m_A$\,[GeV]} &
  \multicolumn{1}{c|}{3000} &
  \multicolumn{1}{c|}{4000} &
  \multicolumn{1}{c|}{4000} &
  \multicolumn{1}{c|}{3000} &
  \multicolumn{1}{c|}{4000} &
  \multicolumn{1}{c|}{4200} \\ \hline
\multicolumn{1}{|c|}{$m_h$\,[GeV]} &
  \multicolumn{1}{c|}{122.7} &
  \multicolumn{1}{c|}{123.3} &
  \multicolumn{1}{c|}{123.3} & 
   \multicolumn{1}{c|}{123.6} &
  \multicolumn{1}{c|}{123.8} &
  \multicolumn{1}{c|}{123.7}
   \\
\multicolumn{1}{|c|}{$m_{\tilde\chi_1^0}$\,[GeV]} &
  \multicolumn{1}{c|}{199.4} &
  \multicolumn{1}{c|}{300.3} &
   \multicolumn{1}{c|}{350.5} &
  \multicolumn{1}{c|}{199.40} &
  \multicolumn{1}{c|}{300.4} &
   \multicolumn{1}{c|}{603.5}
   
   \\ \multicolumn{1}{|c|}{$m_{\tilde\chi_2^0}$\,[GeV]} &
  \multicolumn{1}{c|}{831.9} &
  \multicolumn{1}{c|}{829.5} &
  \multicolumn{1}{c|}{829.1} & 
   \multicolumn{1}{c|}{241.2} &
  \multicolumn{1}{c|}{323.0} &
  \multicolumn{1}{c|}{618.5}
  
   \\\multicolumn{1}{|c|}{$m_{\tilde\chi_1^\pm}$\,[GeV]} &
  \multicolumn{1}{c|}{831.7} &
  \multicolumn{1}{c|}{829.2} &
  \multicolumn{1}{c|}{828.6} & 
   \multicolumn{1}{c|}{245.3} &
  \multicolumn{1}{c|}{323.1} &
  \multicolumn{1}{c|}{618.4}
   \\ 
\multicolumn{1}{|c|}{$m_{\tilde{e}_R}$\,[GeV]} &
  \multicolumn{1}{c|}{206} &
  \multicolumn{1}{c|}{303} &
  \multicolumn{1}{c|}{351} & 
   \multicolumn{1}{c|}{206} &
  \multicolumn{1}{c|}{319} &
  \multicolumn{1}{c|}{606}
   \\
\multicolumn{1}{|c|}{$m_{\tilde{e}_L}$\,[GeV]} &
  \multicolumn{1}{c|}{244} &
  \multicolumn{1}{c|}{363} &
  \multicolumn{1}{c|}{405} & 
   \multicolumn{1}{c|}{239} &
  \multicolumn{1}{c|}{373} &
  \multicolumn{1}{c|}{643}
   \\
\multicolumn{1}{|c|}{$m_{\tilde{\mu}_R}$\,[GeV]} &
  \multicolumn{1}{c|}{210} &
  \multicolumn{1}{c|}{308} &
  \multicolumn{1}{c|}{353} & 
   \multicolumn{1}{c|}{210} &
  \multicolumn{1}{c|}{327} &
  \multicolumn{1}{c|}{617}
   \\
\multicolumn{1}{|c|}{$m_{\tilde{\mu}_L}$\,[GeV]} &
  \multicolumn{1}{c|}{248} &
  \multicolumn{1}{c|}{369} &
  \multicolumn{1}{c|}{407} & 
   \multicolumn{1}{c|}{243} &
  \multicolumn{1}{c|}{380} &
  \multicolumn{1}{c|}{655}\\
\multicolumn{1}{|c|}{$m_{\tilde{\nu}_e}$\,[GeV]} &
  \multicolumn{1}{c|}{231} &
  \multicolumn{1}{c|}{354} &
  \multicolumn{1}{c|}{397} & 
   \multicolumn{1}{c|}{226} &
  \multicolumn{1}{c|}{364} &
  \multicolumn{1}{c|}{638}\\
\multicolumn{1}{|c|}{$m_{\tilde{\nu}_\mu}$\,[GeV]} &
  \multicolumn{1}{c|}{235} &
  \multicolumn{1}{c|}{359} &
  \multicolumn{1}{c|}{399} & 
   \multicolumn{1}{c|}{230} &
  \multicolumn{1}{c|}{371} &
  \multicolumn{1}{c|}{649}
   \\\hline
\multicolumn{1}{|c|}{${\rm Br}(B\rightarrow X_s \gamma)\times 10^4$} &
  \multicolumn{1}{c|}{3.19} &
  \multicolumn{1}{c|}{3.14} &
   \multicolumn{1}{c|}{3.15} &
  \multicolumn{1}{c|}{3.18} &
  \multicolumn{1}{c|}{3.16} &
  \multicolumn{1}{c|}{3.13} \\
\multicolumn{1}{|c|}{${\rm Br}(B^+\to \tau^+\,\nu_\tau)\times 10^{4}$} &
  \multicolumn{1}{c|}{1.25} &
  \multicolumn{1}{c|}{1.24} &
  \multicolumn{1}{c|}{1.24} &
  \multicolumn{1}{c|}{1.24} &
  \multicolumn{1}{c|}{1.24} &
  \multicolumn{1}{c|}{1.24}
   \\
\multicolumn{1}{|c|}{${\rm Br}(B_s\to \mu^+ \mu^-)\times 10^{9}$} &
  \multicolumn{1}{c|}{3.17} &
  \multicolumn{1}{c|}{3.13} &
   \multicolumn{1}{c|}{3.13} &
  \multicolumn{1}{c|}{3.17} &
  \multicolumn{1}{c|}{3.17} &
   \multicolumn{1}{c|}{3.11}\\
\multicolumn{1}{|c|}{$ {\rm Br}(B\to X_s \,\nu\,\overline{\nu})\times 10^{5}$} &
  \multicolumn{1}{c|}{3.98} &
  \multicolumn{1}{c|}{3.98} &
  \multicolumn{1}{c|}{3.98} &
  \multicolumn{1}{c|}{3.98} &
  \multicolumn{1}{c|}{3.98} &
   \multicolumn{1}{c|}{3.98}\\
\multicolumn{1}{|c|}{$a_\mu^{\rm SUSY}$ $\times 10^{9}$} &
  \multicolumn{1}{c|}{1.77} &   
  \multicolumn{1}{c|}{2.14} &
  \multicolumn{1}{c|}{1.59} &
  \multicolumn{1}{c|}{2.82} &
  \multicolumn{1}{c|}{1.94} &
  \multicolumn{1}{c|}{1.51}
   \\ \hline
\multicolumn{1}{|c|}{$\Omega_{\rm DM}h^2$} &
  \multicolumn{1}{c|}{0.119} &
  \multicolumn{1}{c|}{0.122} &
   \multicolumn{1}{c|}{0.121}&
  \multicolumn{1}{c|}{0.121} &
  \multicolumn{1}{c|}{0.119} &
   \multicolumn{1}{c|}{0.118}\\
\multicolumn{1}{|c|}{$\sigma_{\rm SI}^p$\,[pb]$\times 10^{11}$} &
  \multicolumn{1}{c|}{2.40} &
  \multicolumn{1}{c|}{3.77} &
  \multicolumn{1}{c|}{5.46} &
  \multicolumn{1}{c|}{2.95} &
  \multicolumn{1}{c|}{4.16} &
  \multicolumn{1}{c|}{8.88} \\
\multicolumn{1}{|c|}{$\langle \sigma_{\rm ann} v\rangle$\,[cm$^3$/s]$\times 10^{28}$} &
  \multicolumn{1}{c|}{0.95} &
  \multicolumn{1}{c|}{1.19} &
  \multicolumn{1}{c|}{1.30} &
  \multicolumn{1}{c|}{1.08} &
  \multicolumn{1}{c|}{1.31} &
  \multicolumn{1}{c|}{1.73} \\ \hline
\end{tabular}
\caption{A few BMPs that we consider here to illustrate the indirect search prospects of the $\tilde B_{\tilde H}$ and $\tilde B_{\tilde W\tilde H}$ dark matter regimes in the presence of an SMBH. All BMPs yield correct relic abundance and accommodate $\delta a_\mu$ through $a^{\rm SUSY}_\mu$ (Eq.\,\ref{eq:delamu}). All the relevant experimental constraints {\it e.g.} as listed in sec.\,\ref{sec:3} are respected. Apart from the parameters mentioned above, we set the gluino mass, $M_3=4$\,TeV, stau masses $m_{\tilde{\tau}_{L,R}}=1.3$\,TeV, squark soft masses at $5$\,TeV. 
}
\label{tab:BMPs}
\end{table*}

\begin{figure}[ht]
    \centering
    \includegraphics[width=0.6\linewidth]{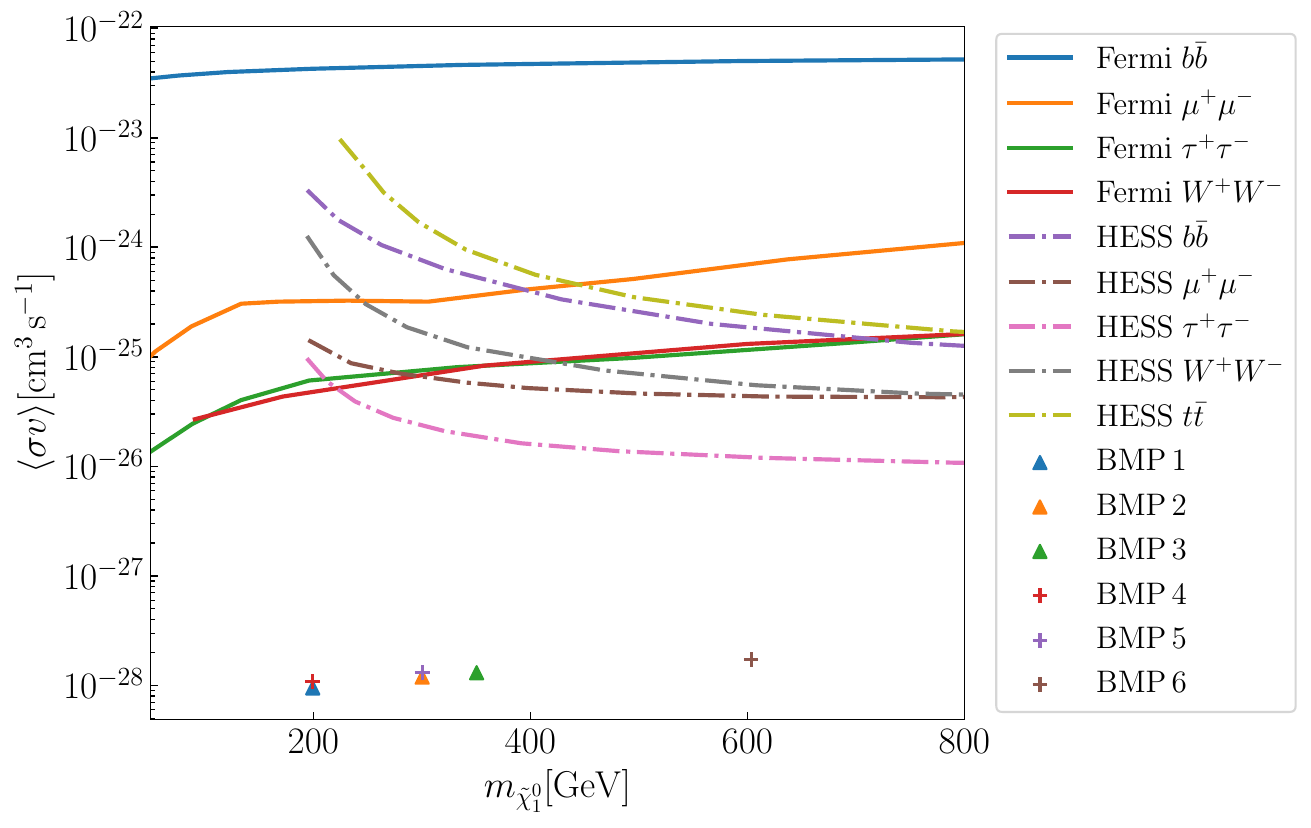}
    \caption{Comparison of total $\langle\sigma v\rangle$ values for BMP\,1\,-\,BMP\,6 with experimental upper limits on DM pair annihilation cross-section to different final state particles. The Fermi-LAT upper bounds (for the NFW profile) to different final state particles assuming $100\%$ branching ratio are from the observations of dSphs \cite{MAGIC:2016xys}, and the indicated HESS bounds (for the Einasto profile) are from observations of GC \cite{HESS:2022ygk}.}
    \label{fig:sigmavcomp}
\end{figure}

Among the chosen points, the co-annihilation with sleptons dominates for BMP\,1\,-\,BMP\,3, which are within the $\tilde B_{\tilde H}$ set. 
The remaining BMPs (BMP\,4\,-\,BMP\,6) come under $\tilde B_{\tilde W \tilde H}$ category, and in this case, co-annihilation with charginos additionally become important. For BMP\,4, the masses of lightest charginos and second lightest neutralino are still far from the DM mass, and thus DM coannihilations with these particles do not contribute significantly to the relic density. Consequently, coannihilations with lighter sleptons remain the dominant contribution for BMP\,4.
In BMP\,5 and BMP\,6, the mass difference between the LSP and charginos is relatively smaller, thus making chargino coannihilations the leading contributing processes. 
 In both sets, the annihilation cross-sections remain relatively suppressed and negligibly contribute to the DM relic abundance. Additionally, the DM annihilation cross-section in the present time is not sufficiently large to yield observable $\gamma$-ray flux from the galactic halo. This is evident from Fig.\,\ref{fig:sigmavcomp}, where we present the total $\langle\sigma v\rangle$ against the available experimental upper limits on the DM pair annihilation cross-sections to different final state particles. 
 
%%%%%%%%%%%%%%%%%%%%%%%%%%%%%%%%%%%%%%%%%%%%%%%%%%%%%%%%%%%%%%%%%%%%
%%%%%%%%%%%%%%%%%%%%%%%%%%%%%%%%%%%%%%%%%%%%%%%%%%%%%%%%%%%%%%%%%%%%%

Before discussing the $\gamma$-ray flux, a few comments on the dominant
decay channels of the EW particles are in order. This will be consistent with the non-observations of any LHC signatures so far. 
For BMPs\,1\,-\,3, ($\tilde{B}_{\tilde{H}}$-case) $\tilde{\chi}_2^0$ and $\tilde{\chi}_3^0$ are dominated by Higgsinos with masses more than 800 GeV and decay to $\tilde{\chi}_1^0$ and $h$ or $Z$ boson. Naturally, the
production of $\tilde\chi_1^\pm \tilde\chi_2^0$ would be suppressed. Similarly,
the direct production of $\tilde e_L^\star \tilde e_L$ or $\tilde \mu_L^\star \tilde \mu_L$ also agrees 
to the necessary mass splitting between $\tilde e_L (\tilde \mu_L) -\tilde\chi_1^0$.
On the other hand, for BMPs\,4\,-\,6, ($\tilde{B}_{\tilde{W} \tilde{H}}$-case ) $\tilde{\chi}_2^0$ is Wino-like with masses very close to $m_{\tilde{\chi}_1^0}$.
Here $\tilde{\chi}_2^0$ decays dominantly to $\nu_{e,\mu}\tilde{\nu}_{e,\mu}$ for BMP\,4 and $e^\pm\tilde{e}_L^\mp$ for BMP\,5 and BMP\,6. 
On the other hand, the lightest chargino mainly decays to an electron, and the left-handed electron sneutrino for BMP\,4. In the case of BMP\,5, the decay favours a three-body process via an off-shell $W$ to $\tilde{\chi}_1^0, u(c), \overline{d}(\overline{s})$. For all the BMPs, $\tilde e, \tilde \mu$ are consistent with the
experimental limits as described in Ref.~\cite{CMS:2018eqb, ATLAS:2022hbt}.

\begin{figure*}[ht]
    \centering
   \includegraphics[width=\textwidth]{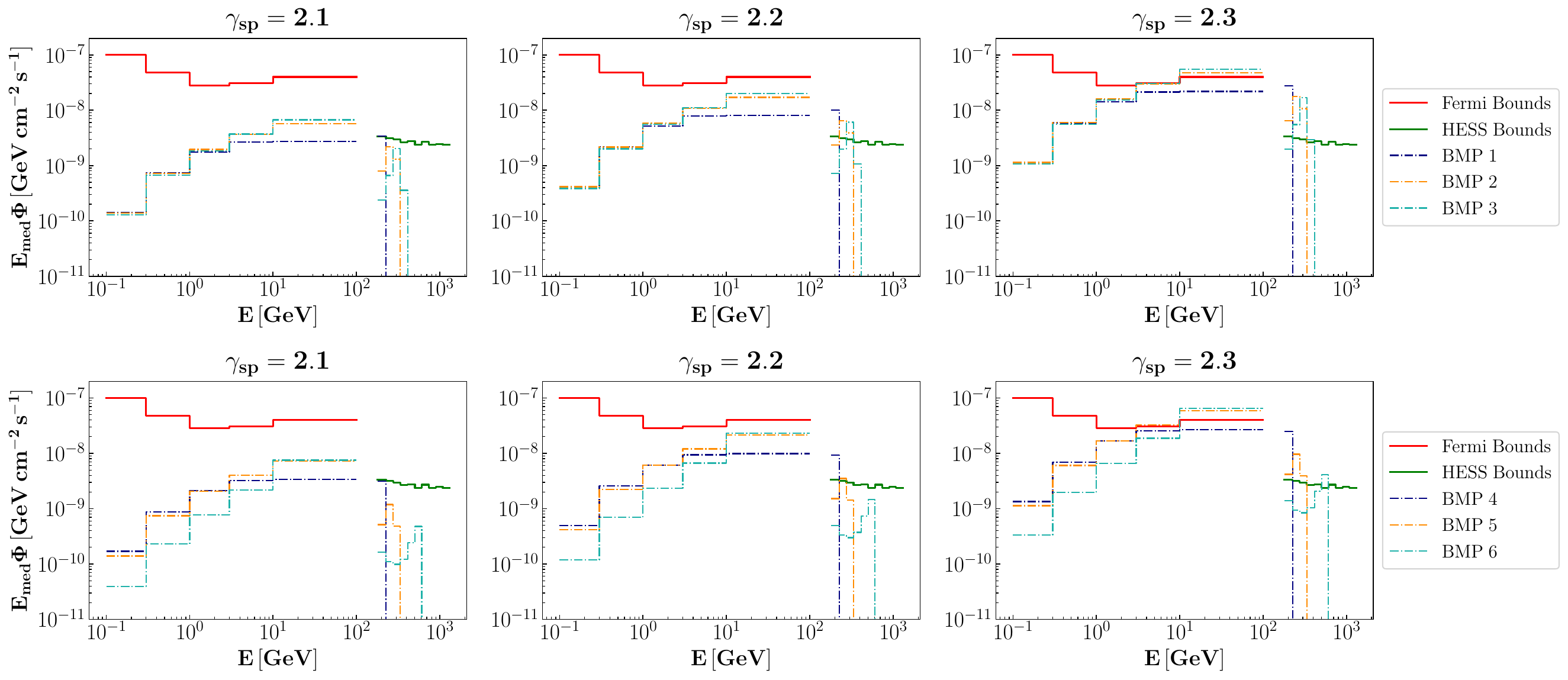}
   \caption{Comparison between predictions of $\gamma$-ray energy flux at each bin due to the density spike around Sgr\,A$^\ast$ for different BMPs mentioned in Table\,\ref{tab:BMPs} and the observed ones from Fermi point source \texttt{3FGL J1745.6\,-\,2859c}\,\cite{Fermi-LAT:2015bhf} and HESS source \texttt{J1745\,-\,290}\,\cite{HESS:2016pst} associated with Sgr\,A$^*$. Different $\gamma_{\rm sp}$ values have been considered with $\gamma_c$ set at 1.2. Here, $E_{\rm med}$ is the central value of the energy of the bin on the logarithmic scale.}
    \label{fig:BMPsGamma}
\end{figure*}

Since the compressed nature of SUSY spectra does not offer the required handle to probe the BMP\,1-6 through collider and DM searches, which otherwise satisfies all other experimental and theoretical constraints, we now like to see whether Milky Way's SMBH Sgr A$^*$ can provide the desired boost in the DM indirect searches to test these specific zones of the $\tilde{B}_{\tilde{H}}$ and $\tilde{B}_{\tilde{W}\tilde{H}}$ regimes. In particular, we estimate the $\gamma$-ray spectra resulting from the DM pair annihilations in the presence of the DM density spike (Eq.(\ref{eq:spike})) for all the six benchmark points listed in Table\,\ref{tab:BMPs}. In Fig.\,\ref{fig:BMPsGamma}, we compare our prediction with the observation of Fermi point source \texttt{3FGL J1745.6\,-\,2859c}\,\cite{Fermi-LAT:2015bhf} within the energy range, $100\,\text{MeV}\leq E_\gamma\leq100\,\text{GeV}$ and HESS source \texttt{J1745\,-\,290}\,\cite{HESS:2016pst} for $180\,\text{GeV}\leq E_\gamma\leq 79\,\text{TeV}$ associated with Sgr\,A$^*$. To compare our results with an available experimental bound, we evaluate the energy flux prediction in each bin used in the corresponding experimental analysis. The photon flux $\Phi_i$ for the $i$-th bin is assessed as:
\begin{align}
(\Phi)_i&=\int_{E_{\rm min}^i}^{E_{\rm max}^i}dE_\gamma\frac{d\Phi_\gamma}{dE_\gamma},
\end{align}
where, $E^i_{\rm max}-E^i_{\rm min}$ is the span of the $i$-th bin. The energy flux of each bin can be obtained by multiplying the total photon flux of the bin to its median energy given by
$(E_{\rm med})_i=\sqrt{E_{\rm min}^iE_{\rm max}^i}$. Thus, the total energy flux for $i$-th bin can be evaluated as:
\begin{align}
    (E_{\rm med}\Phi)_i=\sqrt{E_{\rm min}^iE_{\rm max}^i}\int_{E_{\rm min}^i}^{E_{\rm max}^i}dE_\gamma\frac{d\Phi_\gamma}{dE_\gamma}.
\end{align}

Subsequently, we consider three different $\gamma_{\rm sp}$ values when computing the $\gamma$-ray energy flux for the BMPs in Fig.\,\ref{fig:BMPsGamma}. We notice that for any of the BMPs, the amount of $\gamma$-ray flux increases with the rise of $\gamma_{\rm sp}$, as a relatively larger $\gamma_{\rm sp}$ amplifies the amount of spike in the DM density as indicated in Fig.\,\ref{fig:figSpike}. If, for a particular BMP, the calculated $\gamma$-ray flux exceeds the observed flux from either the Fermi or the HESS data in any of the energy bins, it will be considered as disfavored. For both $\tilde B_{\tilde H}$ and $\tilde B_{\tilde W\tilde H}$ regimes, the choice of $\gamma_{\rm sp}=2.3$ essentially rules out all the BMPs as viable points that can address DM relic abundance and $\delta a_\mu$ simultaneously. 
With a relatively small value for $\gamma_{\rm sp}$, the computed $\gamma$-ray flux values fall below the allowed limits for at least some of the BMPs that we have considered for the $\tilde B_{\tilde H}$ and $\tilde B_{\tilde W\tilde H}$ zones of MSSM.
Thus the above exercise establishes the importance
of DM density spikes 
around Milky Way's SMBH Sgr A$^*$.In particular, it
may effectively provide the necessary boost to the $\gamma$-ray flux observable to 
probe the $\tilde B_{\tilde H}$ and $\tilde B_{\tilde W\tilde H}$ regimes of MSSM. 

\begin{figure}[h]
    \centering
    \includegraphics[width=0.5\linewidth]{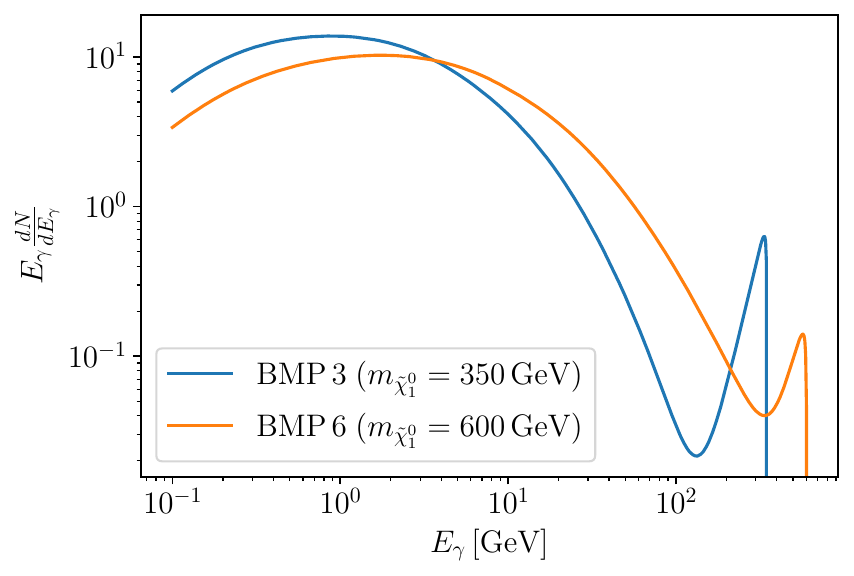}~~
    \caption{Differential photon spectra, $E_\gamma\frac{dN}{dE_\gamma}$ as a function of $E_\gamma$ for BMP\,3 and BMP\,6.}
    \label{fig:mdm_peak}
\end{figure}

The behaviour of photon energy flux $E_{\rm med}\Phi$ in Fig.\,\ref{fig:BMPsGamma} at different energies can be traced to the differential spectra ($dN/dE_\gamma$) of individual annihilation channels (see Eq.(\ref{eq:dndeIndiv})). 
The dominant annihilation modes of the DM for the BMPs at the present epoch are, 
\begin{align}\label{eq:DmannIn}
    \tilde B_{\tilde H}:~~~
    &\text{BMP\,1}:~~~\tilde{\chi}_1^0 \tilde{\chi}_1^0\to \overline{t}t, \,\gamma e^+e^-,\,\gamma\mu^+\mu^-,\\
    &\text{BMP\,2}:~~~\tilde{\chi}_1^0 \tilde{\chi}_1^0\to \overline{t}t, \,\gamma e^+e^-,\,\gamma\mu^+\mu^-,\label{eq:BMP2chan}\\
    &\text{BMP\,3}:~~~\tilde{\chi}_1^0 \tilde{\chi}_1^0\to \overline{t}t, \,\gamma e^+e^-,\,\gamma\mu^+\mu^-,\\
    \tilde B_{\tilde W\tilde H}:~~~
    &\text{BMP\,4}:~~~\tilde{\chi}_1^0 \tilde{\chi}_1^0\to \overline{t}t, \,\gamma e^+e^-,\,\gamma\mu^+\mu^-,\\
    &\text{BMP\,5}:~~~\tilde{\chi}_1^0 \tilde{\chi}_1^0\to \overline{t}t,  \,W^+W^-,\,\gamma e^+e^-,\label{eq:BMP5chan}\\
    &\text{BMP\,6}:~~~\tilde{\chi}_1^0 \tilde{\chi}_1^0\to W^+W^-,\,\overline{t}t,  \,b\overline{b}.
\end{align}
The channels with leptonic final states produce high energy photons with the corresponding $dN_i/dE_\gamma$ peaked at energies close to the DM mass. 
On the other hand, quarks, as final states, produce a copious number of photons during hadronization with relatively lower energies. The spectrum $dN_i/dE_\gamma$ from these channels is expected to exhibit peaks at lower energies \cite{Slatyer:2017sev, Slatyer:2021qgc}. The spectrum of the channels having $W^\pm$ as final states also shows similar behaviour as in the case of hadrons as the produced $W^\pm$ bosons from DM annihilation subsequently decay to quarks with significant branching ratios. The overall differential photon spectrum thus usually features two maxima, one at an energy close to the DM mass ($E_\gamma\simeq m_{\tilde\chi_1^0}$) and another at a relatively lower energy. The appearance of two peaks is evident from Fig.\,\ref{fig:mdm_peak} where we have shown the variation of differential photon spectra $E_\gamma{dN}/{dE_\gamma}$ as a function of $E_\gamma$ for BMP\,3 and BMP\,6.
The amplitude of each peak crucially depends on the branching ratio of DM annihilation modes having quarks and leptons in the final states. 
In the BMPs\,1-3, the leptonic final states ($\gamma e^+e^-$ and $\gamma \mu^+\mu^-$) are among the first three dominantly contributing DM annihilation processes. The dominant presence of 3-body leptonic final states leads to a prominent second peak in the energy range covered by HESS. 

\vfill
%%%%%%%%%%%%%%%%%%%%%%%%%%%%%%%%%%
%%%%%%%%%%%%%%%%%%%%%%%%%%%%%%%%%%%
\vfill

In the BMP\,4, which although belongs to $B_{\tilde{W}\tilde{H}}$ category, $M_2$ is still far from $M_1$. Thus, the Wino component in the DM is still small, which results in the dominant DM annihilation modes and energy flux similar to the $\tilde B_{\tilde H}$ case. On the contrary, in BMP\,5, the dark matter has a relatively larger Wino component, making $\tilde{\chi}_1^0 \tilde{\chi}_1^0 \to W^+W^-$ channel dominating over the ones with leptonic final states. 
In BMP\,6, which has even more significant Wino mixing in DM, the branching ratios for $\tilde{\chi}_1^0 \tilde{\chi}_1^0 \to W^+W^-$ further increases, and the leptonic final states contribute even less. 

In general, a larger Wino component reduces the branching ratio for leptonic final states, leading to suppression of the second peak at the photon flux spectrum, as can be seen in Fig.\,\ref{fig:mdm_peak}. Hence, for the BMPs ({\it, e.g. }BMP\,6) having a mild contribution from leptonic final states, the Fermi-LAT data may also become relevant alongside the HESS data for exclusion as we increase the spike parameter. %\RP{} 
It is important to note that the two-body processes with leptonic final states, \(\tilde{\chi}_1^0 \tilde{\chi}_1^0 \to e^+e^-,\mu^+\mu^-\), are helicity suppressed by typically a factor of $\sim m_{e,\mu}^2/m^2_{\tilde\chi_1^0}$ since the DM is a non-relativistic Majorana particle \cite{Bringmann:2007nk, Bergstrom:1989jr, Goldberg:1983nd}, thereby allowing the three-body processes to contribute predominantly to the photon flux at \(E_\gamma \simeq m_{\tilde{\chi}_1^0}\).

\vspace{3mm}

\begin{figure*}[h]
    \centering
   \includegraphics[width=0.7\textwidth]{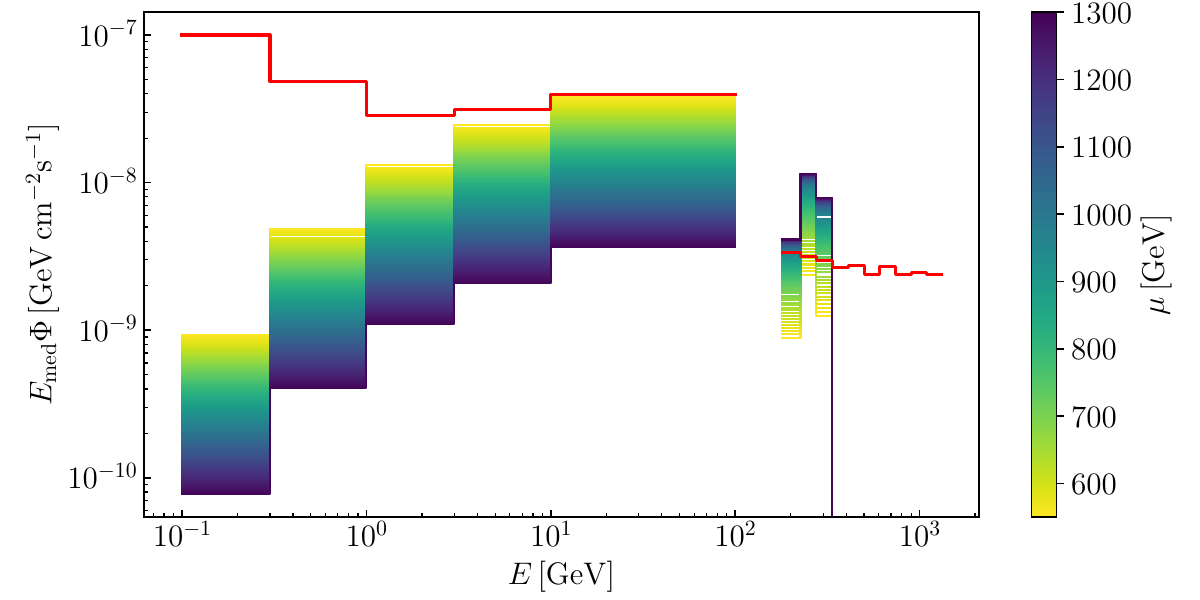}\\
    \includegraphics[width=0.424\textwidth]{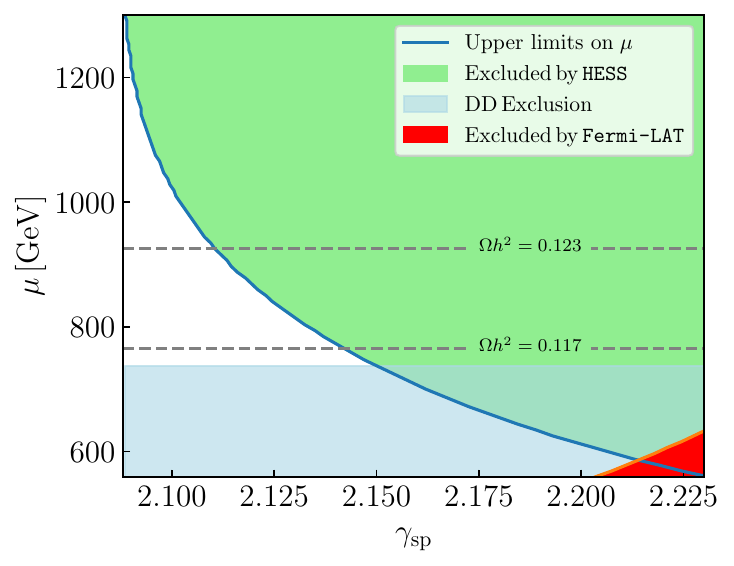}
~~~~~~  \includegraphics[width=0.49\textwidth]{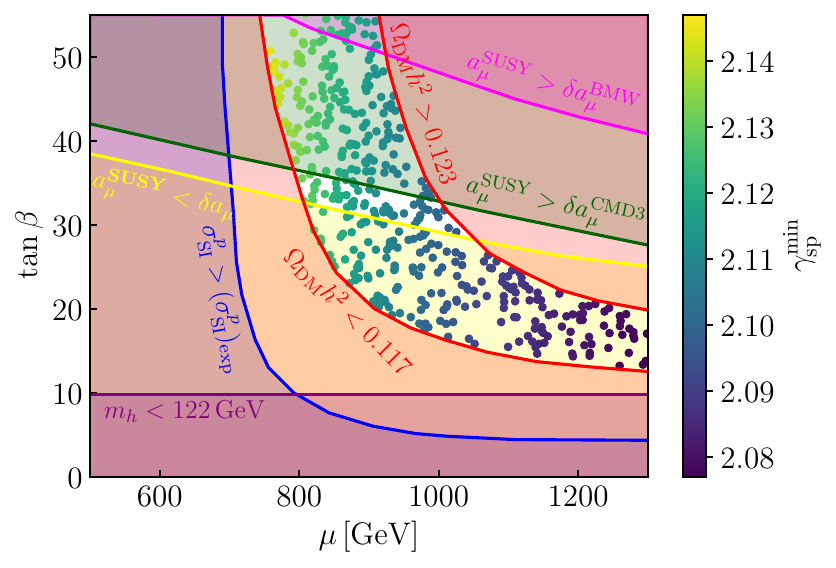}
   \caption{Results for a $\tilde B_{\tilde H}$ scenario. Top panel: Variation of $\gamma$-ray energy flux with energy for varying $\mu$ considering $\gamma_{\rm sp}=2.2$.  Except $\mu$, other relevant parameters are set as in BMP\,2 of Table\,\ref{tab:BMPs}. Bottom left panel: Excluded (coloured) and allowed regions (white {region between the dashed lines})  in the $\gamma_{\rm sp}\,$-$\,\mu$ plane considering relic density, SI-DD, HESS and Fermi-LAT data. Bottom right panel:  Results of varying $\tan\beta$ and $\mu$ 
   with other parameters set as in BMP\,2.  Excluded regions are shown in solid colours. 
  The scattered circles are graded with the colours of their 
   $\gamma_{\rm sp}^{\rm min}$ values.
   The $\gamma_{\rm sp}^{\rm min}$ values are spread over a small range. Additionally, as depicted, the same circles satisfy constraints due to relic density, SI-DD, $\delta a_\mu$ (via Eq. \eqref{eq:delamu}), and all the other experimental constraints mentioned earlier.
   In constraining $a_\mu^{\rm SUSY}$, we additionally use the recent BMW 
   (Eq.\eqref{eq:bmw}) and the CMD3 results
   (Eq.\eqref{eq:cmd3}) with $2\sigma$  C.L. For the latter choices, the parameter space points with 
   low $\tan\beta$ are still viable. Similarly, heavier neutral and charged Higgsinos alongside $\tilde \mu$ are now allowed.
   Thus, each viable point with a $\gamma_{\rm sp}$ value smaller than $\gamma_{\rm sp}^{\rm min}$ would be allowed by both Fermi-LAT and HESS data. This, as well as all the subsequent figures, refer to $\gamma_c=1.2$\,.}
  
    \label{fig:BMP1mu}
\end{figure*}

\begin{figure*}[ht]
    \centering
   \includegraphics[width=0.7\textwidth]{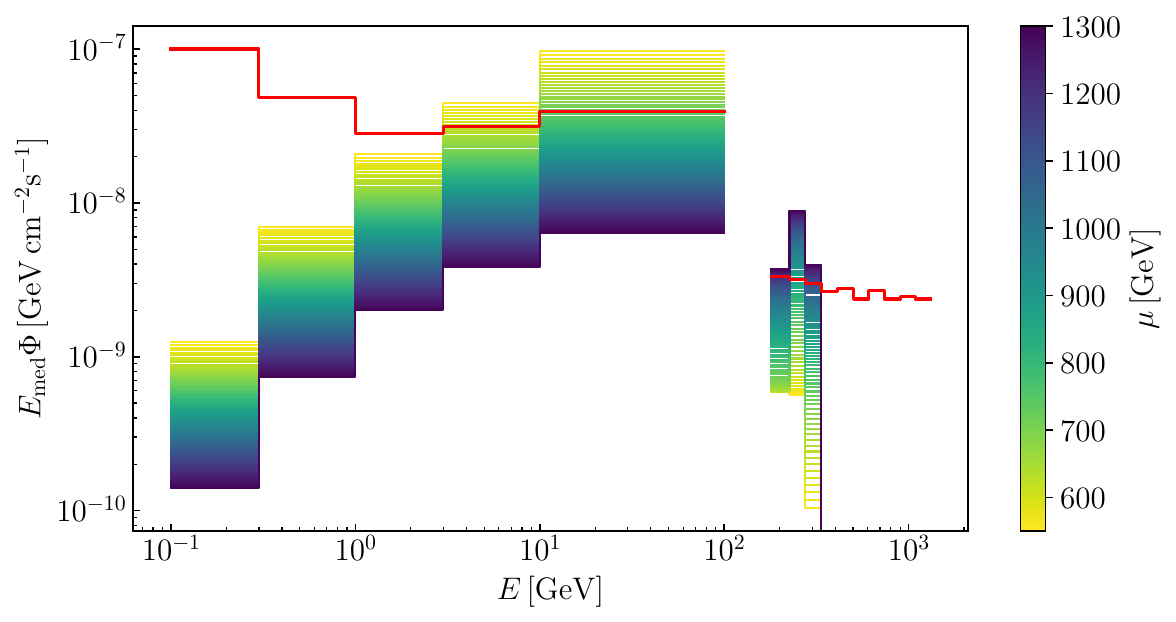}\\
\includegraphics[width=0.424\textwidth]{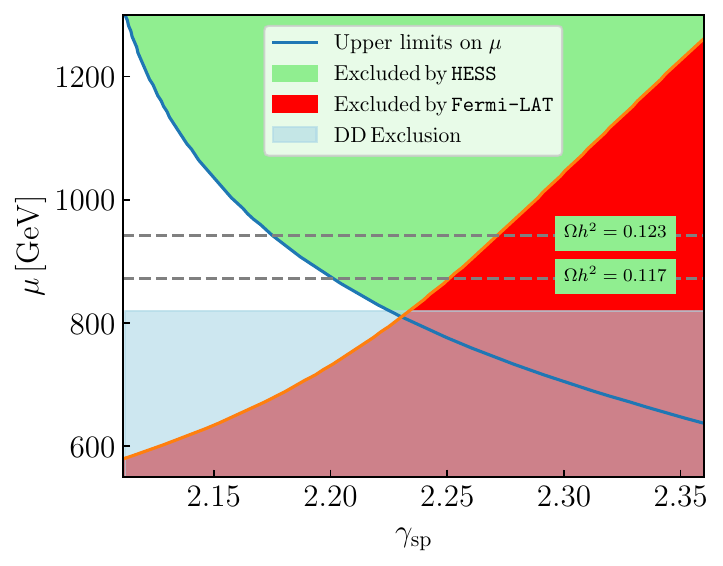}
~~~~~~\includegraphics[width=0.49\textwidth]{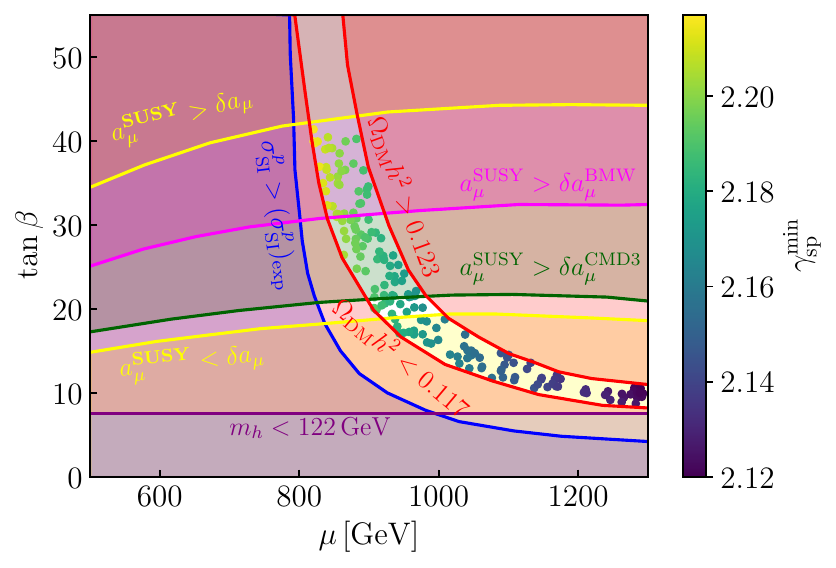}
   \caption{Same as in Fig.\,\ref{fig:BMP1mu}, but for depicting the $\gamma$-ray flux signatures of $\tilde B_{\tilde W\tilde H}$ regime. 
   Here, the input parameters that are held fixed refer to BMP\,5.} 
    \label{fig:BMP4mu}
\end{figure*}

We will now study the parametric dependence of $\Phi_\gamma$ on the choice of the different MSSM and astrophysical inputs. Again, we focus on the quantity $(E_{\rm med}\Phi)_i$, defined earlier and consider both $\tilde B_{\tilde H}$ and 
$\tilde B_{\tilde W\tilde H}$ regimes in this regard.
To begin with, we demonstrate the dependency of photon flux on the $\mu$ parameter in the top panel of Fig.\,\ref{fig:BMP1mu} in the context of $\tilde B_{\tilde H}$ DM. All the input parameters are set in accordance with BMP\,2 
except for the parameters which are varied \footnote{The choice of BMP\,2 and BMP\,5 (which will subsequently be presented to illustrate the parametric dependence of $\gamma$-ray flux in the $\tilde B_{\tilde W \tilde H}$ region in Fig.\,\ref{fig:BMP4mu})
is motivated so as to allow relative comparisons for photon flux at a given $m_{\tilde\chi_1^0}$.}.
In Fig.\,\ref{fig:BMP1mu}, it is apparent that the quantity  $E_{\rm med}\Phi$ rises as $\mu$ decreases in the energy range probed by Fermi. On the contrary, the behaviour is opposite for the energy range relevant to the HESS experiment, where $E_{\rm med}\Phi$ gets reduced as we decrease $\mu$. The behaviour, seemingly quite intriguing, can be traced once again to the differential gamma-ray yield from the spectra of individual annihilation channels as noted in Eq.(\ref{eq:dndeIndiv})\,\footnote{The other inputs, e.g., the $J$-factor (see
Eq.\ref{eq:dPhidE}) does not show any meaningful effect with variations in $\mu$.}. The dominant DM annihilation channels of DM in this case remain similar as specified in Eq.(\ref{eq:BMP2chan}). 
A larger $\mu$ leads to a smaller Higgsino mixing in the DM, thereby reducing the branching ratios of processes with hadronic final states and increasing the contribution of modes with three body leptonic final states. Consequently, the peak at energies near the dark matter mass is enhanced, while the peak at relatively lower energies is diminished.

%%%%%%%%%%%%%%%%%%%%%%%%%%%%%%%%%%%%%%%%%%%%%%
%%%%%%%%%%%%%%%%%%%%%%%%%%%%%%%%%%%%%%%%%%%%%%%%%

The distinct behaviour exhibited by $E_{\rm med}\Phi$ with respect to $E_\gamma$ (top panel of Fig.\,\ref{fig:BMP1mu}) renders the HESS data particularly relevant for higher $\mu$ values which cause the enhancement of the $\gamma$-ray flux at $E_\gamma\geq 180\,{\rm GeV}$. 
On the contrary, Fermi data is more likely to impose constraints on lower $\mu$ values for a given $\gamma_{\rm sp}$. 
The above is evident in the bottom-left panel of Fig.\,\ref{fig:BMP1mu} that identifies allowed and disallowed regions in the plane of $\mu- \gamma_{\rm sp}$ based on the HESS data.  The zones are separated by a solid blue line. Thus, for a given value of $\mu$, the excluded region by HESS data falls on the right of the blue line. The above corresponds to the minimum possible value of $\gamma_{\rm sp}$ (=$\gamma_{\rm sp}^{\rm min}$) above which one obtains a HESS-disallowed zone.
Notably, a higher $\mu$ necessitates a lower $\gamma_{\rm sp}^{\rm min}$ so as to avoid HESS exclusion (see Fig.\,\ref{fig:BMP1mu}). 
However, as shown in Fig.\,\ref{fig:BMP1mu}, SI-DD is more restrictive and 
rules out the parameter space for $\mu\lesssim 720$\,GeV. Furthermore, the observed DM relic density (via Eq.(\ref{eq:relic})) constrains $\mu$ to vary within a narrow range. This permissible range of $\mu$ is expected to shift if one of the other pertinent parameters, such as $\tan\beta$, deviates from the value as in BMP\,2. The figure also shows a region discarded by Fermi-LAT data that appears for small $\mu$ and large values of $\gamma_{\rm sp}$. In the bottom-right panel of Fig.\,\ref{fig:BMP1mu}, we depict the allowed parameter points as coloured circles in the $\mu$-$\tan\beta$ plane consistent with the observed DM relic density, 
$\delta a_\mu$ (via Eq.\eqref{eq:delamu}), and the SM-like Higgs mass while simultaneously satisfying the SI-DD constraint on $\tilde\chi_1^0$ DM. In constraining $a_\mu^{\rm SUSY}$, we additionally use the recent BMW 
   (Eq.\eqref{eq:bmw}) and the CMD3 results 
   (Eq.\eqref{eq:cmd3}) with $2\sigma$ {\it C.L.}\,. For the latter choices, the parameter space points with 
   low $\tan\beta$ are still viable. Similarly, heavier neutral and charged Higgsinos alongside $\tilde \mu$ are now allowed. With the rise in the $\mu$-parameter, DM annihilation to leptonic final states will increase, resulting in a larger flux in the HESS region. Thus, a lower value of $\gamma_{\rm sp}^{\rm min}$ is required\footnote{The BMW and CMD3 results for SM calculations causes the lower limit on ${\mu}$ to increase by $\lesssim20\%$. Thus, the electroweak fine-tuning parameter is expected to increase rather moderately.}.
The scattered circles are graded with colours for their $\gamma_{\rm sp}^{\rm min}$ values.
As evident, the $\gamma_{\rm sp}^{\rm min}$ values are 
spread in a small range, irrespective of $\tan\beta$.

%%%%%%%%%%%%%%%%%%%%%%%%%%%%%%%%%%%%%%%%%%%%%%%%%%%%%%%%%%%%%%%
%%%%%%%%%%%%%%%%%%%%%%%%%%%%%%%%%%%%%%%%%%%%%%%%%%%%%%%%%%%%%%%%%

%%%%%%%%%%%%%%%%%%%%%%%%%%%%%%%%%%%%%%%%%%%%%%%%%%%%%%%%%%%%%%%%%%
\begin{figure*}[h]
    \centering
     \includegraphics[width=0.7\textwidth]{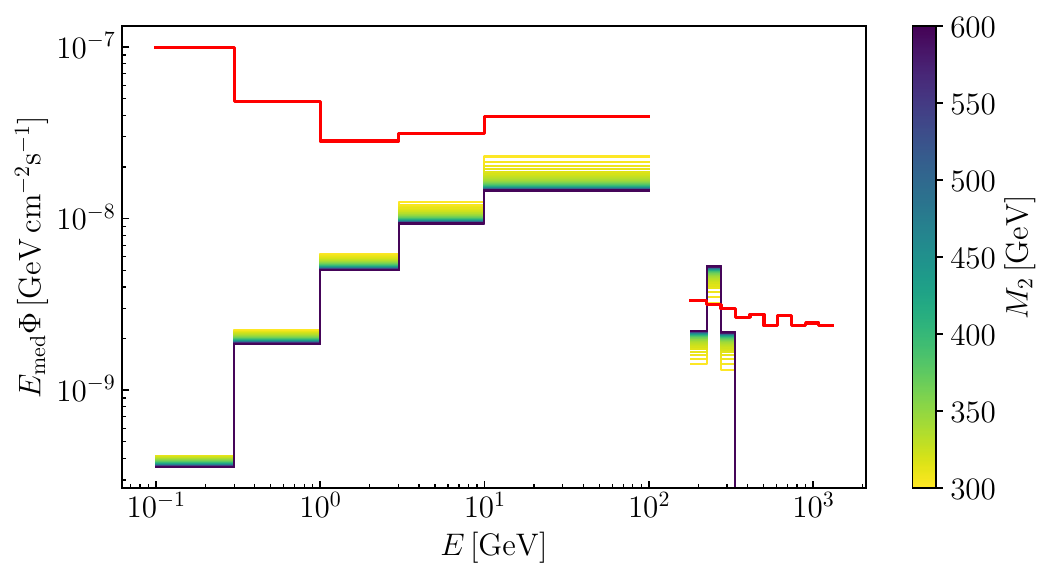}\\
  \includegraphics[width=0.4\textwidth]{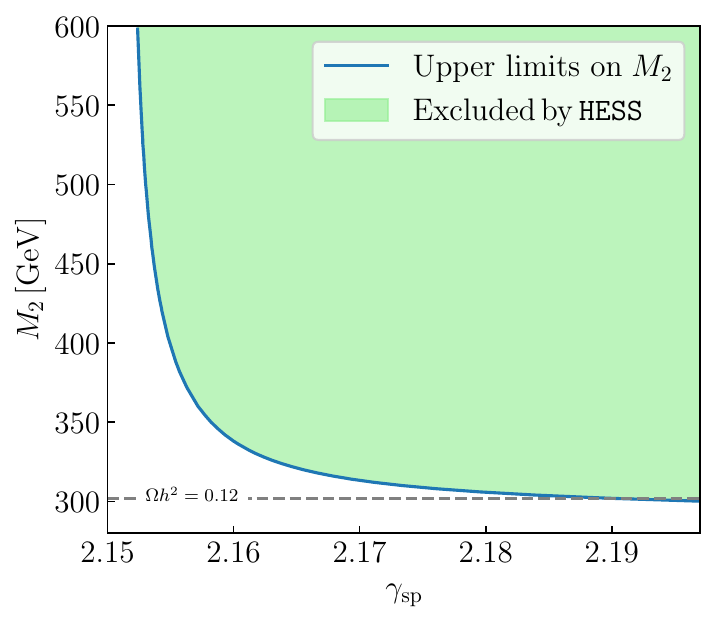}
~~~~~~ \includegraphics[width=0.4\textwidth]{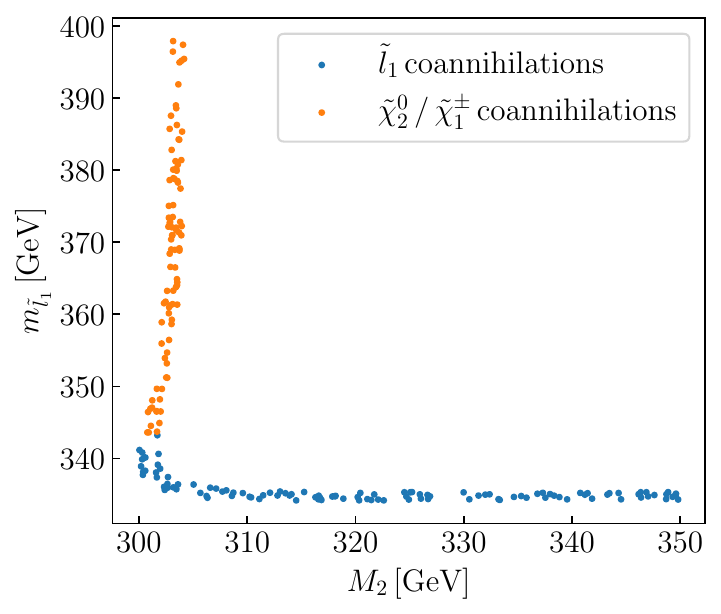}
   \caption{Variation of $\gamma$-ray energy flux with $M_2$ is highlighted in top panel, considering $\gamma_{\rm sp}=2.2$ while the other relevant parameters are set as in BMP\,5 of Table\,\ref{tab:BMPs}. In the bottom left panel, the HESS excluded region is shown in $\gamma_{\rm sp}\,$-$\,M_2$ plane. In the bottom right panel, we consider a deviation of $m_{\tilde{l}_1}$ from the one in BMP\,5 and highlight the viable points in the $M_2$-$m_{\tilde{l}_1}$ plane that provides the desired value of DM relic and $\delta a_\mu$ and also consistent with all other relevant experimental constraints.}
    \label{fig:BMP4muM2}
\end{figure*}

We also perform a comparable examination for the $\tilde B_{\tilde W\tilde H}$ scenario, as depicted in Fig.\,\ref{fig:BMP4mu}, showing how $\mu$ correlates with the photon energy flux. 
Here, we adhere to BMP\,5 for choosing the input parameters except for those under variations. The results presented in Fig.\,\ref{fig:BMP4mu} echo similar observations discussed earlier in Fig.\,\ref{fig:BMP1mu}. 
Noticeably, the Fermi-LAT exclusion zone in this case is larger than the $\tilde B_{\tilde H}$ case. This is due to $M_2$ being close to $M_1$ in BMP\,5, which, as explained earlier, enhances the flux in the lower energy range relevant to Fermi-LAT data.

Focusing on the same $\tilde B_{\tilde W\tilde H}$ scenario, we now explore how a variation of $M_2$ would affect the dark matter sector as shown in Fig.\,\ref{fig:BMP4muM2}.
$M_2$ 
influences the DM relic abundance and $E_{\rm med}\Phi$ at any energy bin.
Reducing $M_2$ 
would lead to a 
rise in the Wino component in the LSP, which in turn would lower the relic abundance via Bino-Wino coannihilations. 
The Higgsino mixing parameter $\mu$ set at 900 GeV (BMP\,5) certainly has a milder influence on the relic density.
However, regarding photon flux, the behaviour
for varying $M_2$ and $\mu$ as evident from the top panels of Fig.\,\ref{fig:BMP4muM2} and Fig.\,\ref{fig:BMP4mu} are not so different.
The dominant annihilation modes of the DM at the present epoch are given in Eq.(\ref{eq:BMP5chan}). Once again, the HESS data emerges as the pertinent factor in constraining $M_2$, especially for the higher values of the latter. 
The bottom left panel of Fig.\,\ref{fig:BMP4muM2} shows that one requires larger $\gamma_{\rm sp}$ that would likely eliminate the smaller $M_2$ region by the HESS data. Both the figures (i.e. the top and the bottom-left panels) affirm that for smaller $M_2$ values, it becomes easier to evade the HESS constraint, necessitating a higher value of $\gamma_{\rm sp}^{\rm min}$ for the exclusion.\footnote{Since the LSP, $\tilde\chi_1^0$, is predominantly a Bino with only a little admixture of Higgsino and Wino,
$\tan\beta$ and slepton masses only have negligible impacts on the photon flux, so we refrain from considering them further.}
The bottom-right panel of Fig.\,\ref{fig:BMP4muM2} shows the relic density satisfied regions arising out of Bino-slepton or Bino-Wino coannihilations.
For our choice of parameters, the relic density is satisfied via Bino-Wino coannihilations in a nearly 
vertical strip around $M_2 \sim 300~$GeV.
As Wino has a higher isospin
coupling compared to Higgsino, we find that varying $M_2$ will cause a significant 
deviation in the relic density. However, 
an enlarged parameter space accommodating the correct amount of DM relic and the required $\delta a_\mu$ (via Eq.\eqref{eq:delamu}) can be generated by varying one of the slepton masses, such as $m_{\tilde\mu}$ 
(or $m_{\tilde e}$) and $M_2$, around their respective values for BMP\,5. 
When $M_2$ is large, reducing $m_{\tilde{\mu}}$ 
(and/or $m_{\tilde e}$) can assist in achieving the correct DM relic by boosting the 
co-annihilations involving slepton. 
On the other hand, the parameter space corresponding to Fig.\,\ref{fig:BMP4muM2} is entirely compliant with SI-DD limits. This is because of a reduced Higgsino fraction in the LSP arising out 
of an adequately large $\mu$ relative to $M_1$.

Guided by the previous lessons, considering both $\tilde B_{\tilde H}$ and $\tilde B_{\tilde W\tilde H}$ scenarios, we now proceed for random scans of the relevant parameters in the specified ranges as listed below:

\vspace{2mm}
\noindent $\bullet$ {\bf Bino-Higgsino} :
Here we fix $M_2=1.5~$TeV. The other relevant parameters are varied in the following ranges.
\begin{equation}
    ~\begin{array}{r@{\,}r@{\,}c@{\,}ll}
 100 \,{\rm GeV}&\leq& ~M_1~ &\leq& 700 \,{\rm GeV},\nonumber\\
 2.0 \,{\rm TeV}&\leq& ~M_A~ &\leq& 4.5 \,{\rm TeV},\\
  500\, {\rm GeV}&\leq& ~\mu~&\leq& 1500\, {\rm GeV},\nonumber\\
  100\,{\rm GeV}&\leq& ~m_{\tilde{l}_{L,R}}~&\leq& 1\,{\rm TeV},\nonumber\\
 5&~\leq& ~\tan\beta~&\leq& 55.%,
\end{array}
\end{equation}
Apart from the parameters mentioned above, we set the gluino mass,
$M_3=4$\,TeV, $m_{\tilde{\tau}_{L,R}}=1.3$\,TeV, squark soft masses at $5$\,TeV and all the trilinear soft SUSY breaking parameters $T_i=-2$ TeV. 
In the above, we have assumed $\tilde l=\{\tilde e,\tilde\mu\}$. During the scan, we have omitted the points for which the lightest neutralino carries a predominant fraction of Higgsino.

\begin{figure*}[h]
    \centering
   \includegraphics[width=0.49\textwidth]{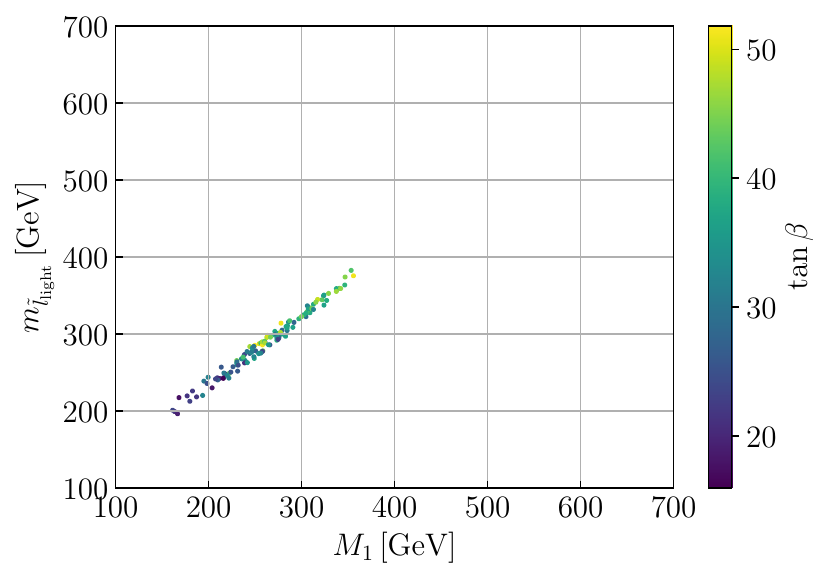}
    \includegraphics[width=0.49\textwidth]{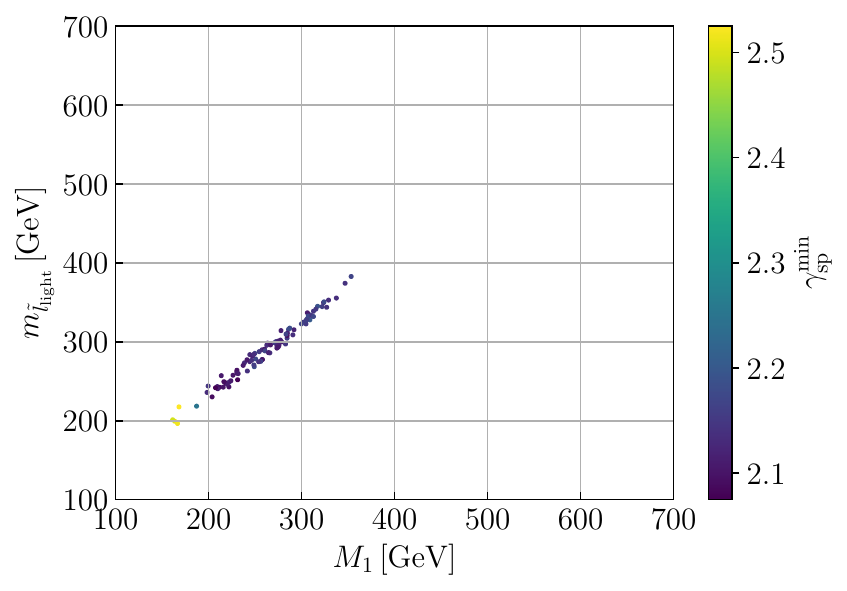}
   \includegraphics[width=0.49\textwidth]{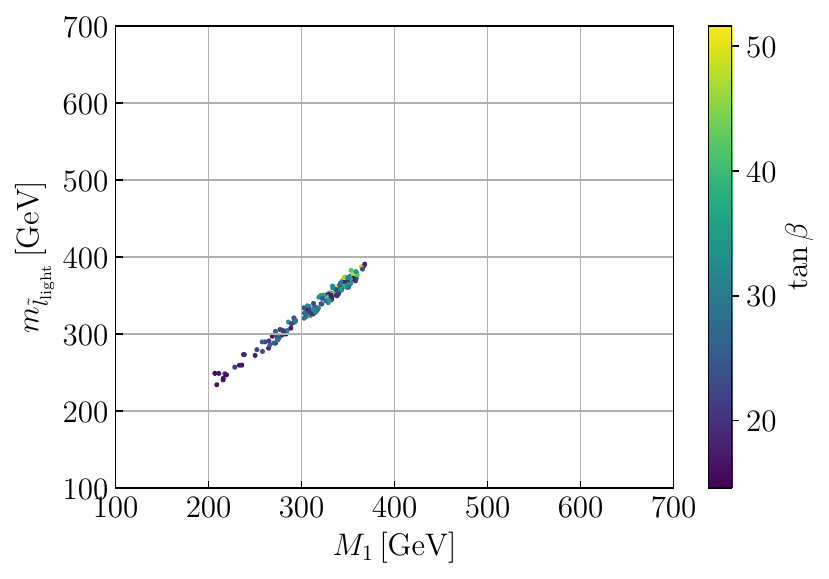}
    \includegraphics[width=0.49\textwidth]{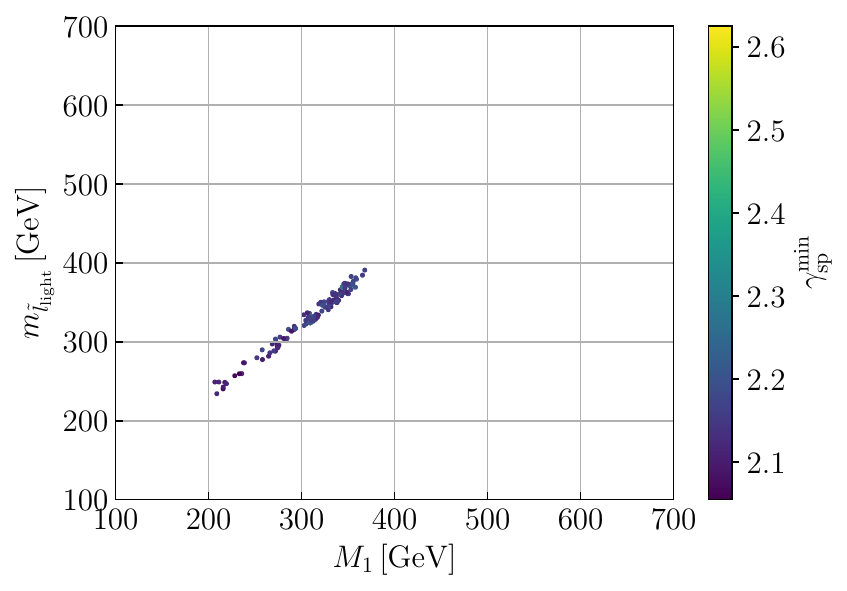}
   \caption{Viable points out of random scanning for the $\tilde B_{\tilde H}$ scenario. However, the points in the top panels consider $a^{\rm SUSY}_\mu \equiv \delta a_\mu$ assuming $5.1\sigma$ discrepancy between SM and experimental average, while the bottom panels follow the CMD3 result which is more stringent than the BMW.
   Left panel: Scatter plots in the $M_1-m_{\tilde{e}_{L,R}}$ plane with points represented by colour-graded circles for $\tan\beta$, the latter shown as a colour bar.  Right panel: Scatter plots in the same plane with points shown as colour-graded circles for $\gamma_{\rm sp}^{\rm min} $, the latter shown as a colour bar. 
   }
    \label{fig:BHRan}
\end{figure*}
%%%%%%%%%%%%%%%%%%%%%%%%%%%%%%%%%%%%%%%%%%

The results are shown in Fig.\,\ref{fig:BHRan}. The left panels show valid parameter points where DM relic density is satisfied via Bino-slepton coannihilations for varying $\tan\beta$ shown in the colour bar.
In addition to the relic density limits, the coloured points 
satisfy the $\delta a_\mu$ data as well as all other relevant constraints as detailed in section\,\ref{sec:3}. 
In the right panels, we have provided the guiding parameter $\gamma_{\rm sp}^{\rm min}$ so that for $\gamma_{\rm sp}<\gamma_{\rm sp}^{\rm min}$ one finds the valid parameter
points allowed by
both the Fermi-LAT and the HESS data. 
Notably, in the $\tilde B_{\tilde H}$ case, the neutralino mass is strictly restricted in the regime: $120\,{\rm GeV}\lesssim m_{\tilde{\chi}_1^0}\lesssim 360\,{\rm GeV}$ beyond which it produces an overabundance of DM.
The irregularity in the
behaviour of $\gamma_{\rm sp}^{\rm min}$ 
in the range 100\,GeV $\lesssim M_1\lesssim$ 200\,GeV is due to the 80 GeV gap between the observed photon energy ranges of Fermi-LAT ($E_\gamma\le100\,\rm GeV$) and HESS ($E_\gamma\ge180\,\rm GeV$) experiments. 
For DM mass in this energy gap, the peak in photon energy flux, which lies at $E_\gamma\sim m_{\tilde\chi_1^0}$, can not be probed by either experiment, requiring a larger $\gamma_{\rm sp}$ value to exclude the parameter set. The bottom panel illustrates the points based on the CMD3 result for $a^{\rm SM}_\mu$. Here, the sleptons, particularly $\tilde \mu$, receive a lower bound due to its large contribution to $a_\mu$. Since the lightest neutralino ($\tilde\chi_1^0$) must form a compressed spectrum with the sleptons to evade LHC constraints, the dark matter mass is limited to values above 200\,GeV.

\vspace{2mm}
\noindent $\bullet$ {\bf Bino-Wino-Higgsino} : Here $M_2$ has also been varied. The ranges of the variations are the same as before, except for the parameter $\mu$ that can now assume relatively higher values.
\begin{align}
    ~\begin{array}{r@{\,}r@{\,}c@{\,}ll}
100 \,{\rm GeV}&\leq& M_1 &\leq& 700 \,{\rm GeV},\nonumber\\
100 \,{\rm GeV} &\leq& M_2 &\leq& 1 \,{\rm TeV},\nonumber\\
 2.0 \,{\rm TeV}&\leq& M_A &\leq& 4.5 \,{\rm TeV},\nonumber\\
 500\, {\rm GeV}&\leq& \mu&\leq&  2.0\, {\rm TeV},\nonumber\\
 100\,{\rm GeV}&\leq& m_{\tilde{l}_{L,R}}&\leq& 1\,{\rm TeV}.\nonumber\\
 5&~\leq& ~\tan\beta~&\leq& 55,
 \end{array}
\end{align}

%%%%%%%%%%%%%%%%%%%%%%%%%%%%%%%%%%%%%%%%%%
\begin{figure*}[h]
    \centering
   \includegraphics[width=0.49\textwidth]{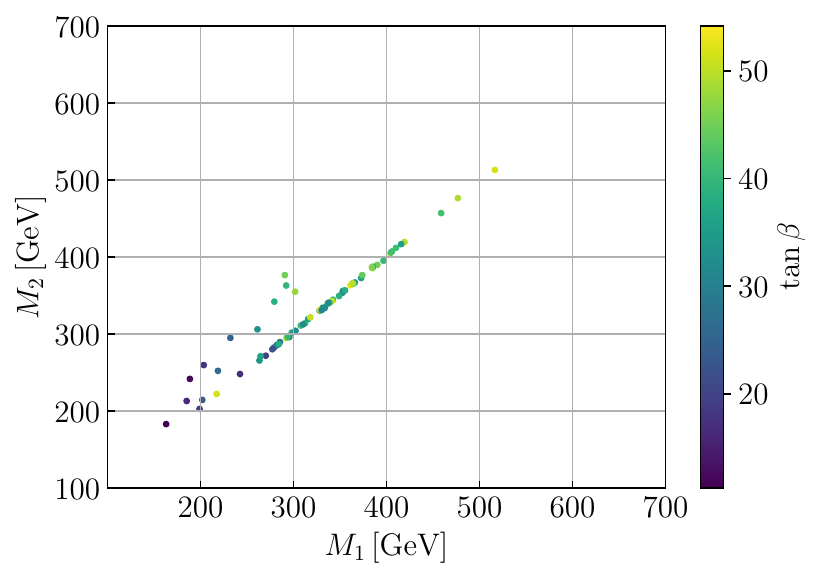}
   \includegraphics[width=0.49\textwidth]{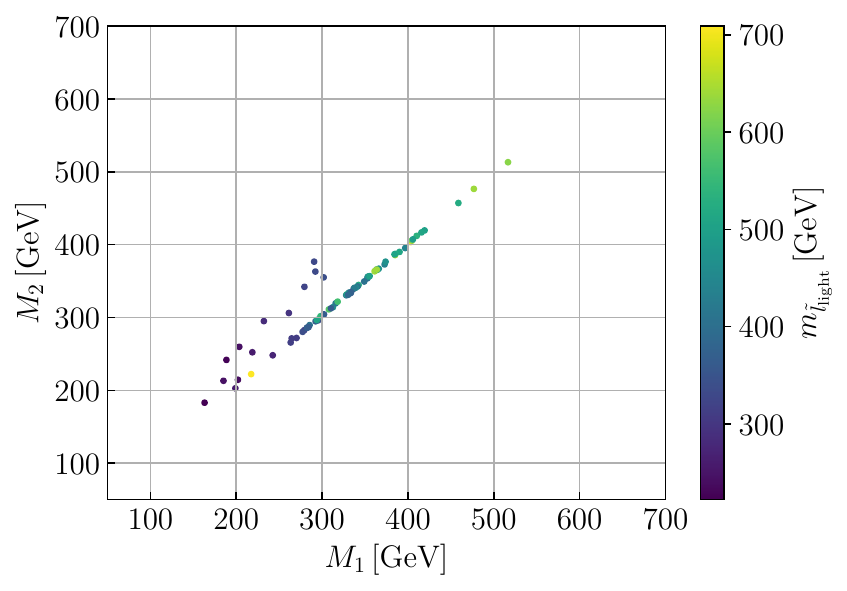}
    \includegraphics[width=0.49\textwidth]{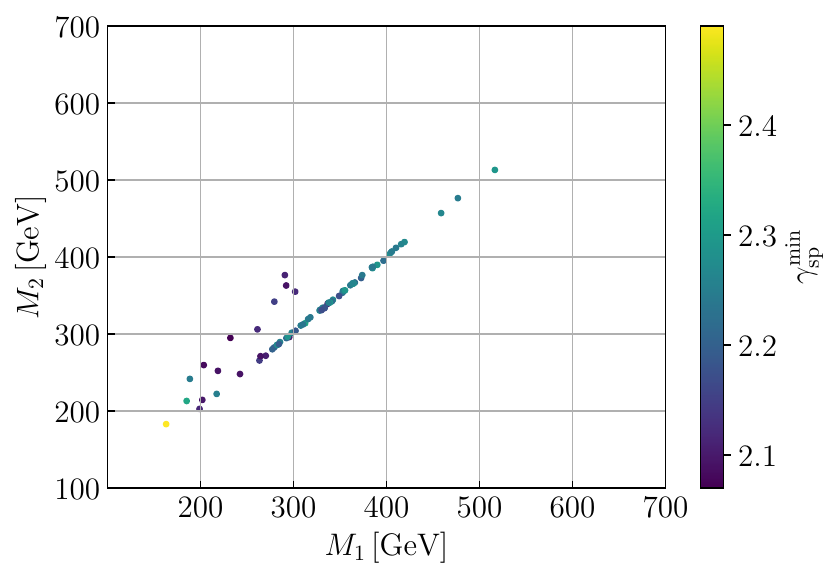}
   \caption{
   Viable points out of random scanning for the $\tilde B_{\tilde W \tilde H}$ scenario. Here
   $a^{\rm SUSY}_\mu \equiv \delta a_\mu$ assuming $5.1\sigma$ discrepancy between SM and experimental average.
   Top left panel:
Scatter plot in the  $M_1$\,-\,$M_2$ plane with points represented by color graded circles for
 $\tan\beta$,  the latter shown as a colour bar. Top right panel: Scatter plot in the same plane with points represented by colour-graded circles for  $m_{\tilde{l}_1}$,  the latter shown as a colour bar. Bottom panel: Scatter plot in the same plane with colour-graded circles for $\gamma_{\rm sp}^{\rm min} $, the latter shown as a colour bar.}
    \label{fig:BWHRan}
\end{figure*}
%%%%%%%%%%%%%%%%%%%%%%%%%%%%%%%%%%%

%%%%%%%%%%%%%%%%%%%%%%%%%%%%%%%%%%%%%%%%%%
\begin{figure*}[h]
    \centering
   \includegraphics[width=0.49\textwidth]{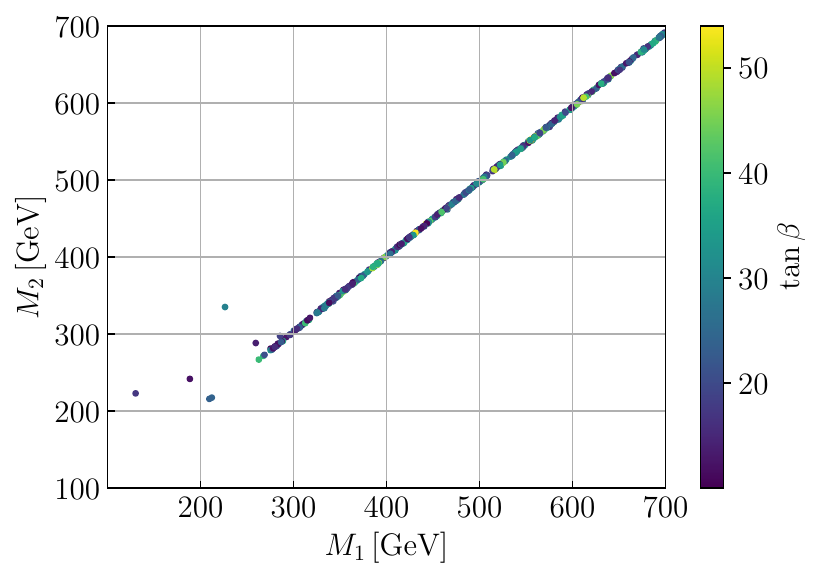}
   \includegraphics[width=0.49\textwidth]{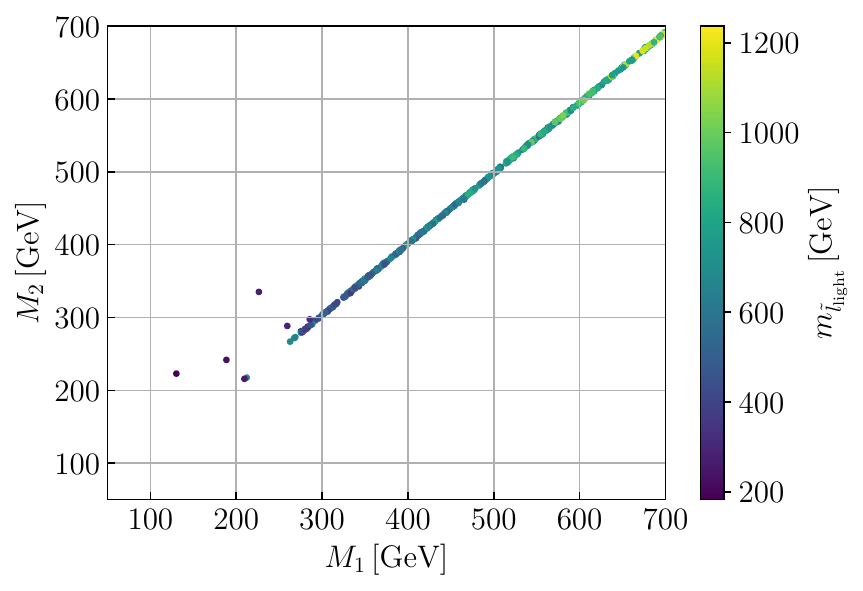}
    \includegraphics[width=0.49\textwidth]{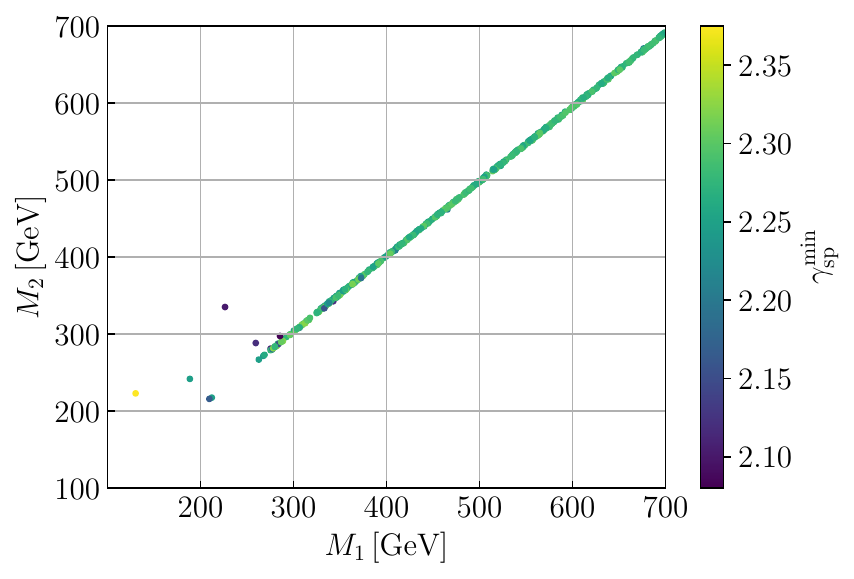}
   \caption{Same as Fig.\,\ref{fig:BWHRan} but $a_\mu^{\rm SUSY}$ is restricted through $\delta a_\mu^{\rm CMD3}$.}
    \label{fig:BWHRanCMD3}
\end{figure*}
%%%%%%%%%%%%%%%%%%%%%%%%%%%%%%%%%%%

All the other parameters are the same as in the previous case. 
Again, we have 
filtered out the parameter points associated with Bino-dominated LSP only while discarding other possibilities. 
The result for $\tilde B_{\tilde W \tilde H}$ case is depicted in Fig.\,\ref{fig:BWHRan}. Here, additional co-annihilation processes involving $\tilde{\chi}_2^0$ and charginos are present, as mentioned earlier. 
 This introduces more flexibility to the MSSM parameter space to achieve the correct value of DM relic abundance. Unlike the $\tilde B_{\tilde H}$ case, the close proximity of $m_{\tilde{l}_{L,R}}$ and $M_1$ is no longer an absolute necessity for satisfying the relic abundance. 
Indeed, the relative variations between $M_2$ and $m_{\tilde{l}_{L,R}}$ result in a relatively broader space where relic abundance can be satisfied in the $M_1$ and $M_2$ plane,
as illustrated in Fig.\,\ref{fig:BWHRan} assuming $5.1\sigma$ discrepancy in $(g-2)_\mu$. We also find a relatively broader mass window for neutralino DM ($120\,{\rm GeV}\lesssim m_{\tilde{\chi}_1^0}\lesssim 530\,{\rm GeV}$) in this case. The requisite value of $\gamma_{\rm sp}^{\rm min}$ value for each feasible point can be found from the bottom panel of Fig.\,\ref{fig:BWHRan}. The irregularity in the range 100\,GeV $\lesssim M_1\lesssim$ 180\,GeV is again due to the gap in observation energies of the experiments, as explained earlier. Fig.\,\ref{fig:BWHRanCMD3} assumes CMD3 results for $a^{\rm SM}_\mu$. Heavier sleptons are now allowed since the difference between
$a_\mu^{\rm SM}$ and $a_\mu^{\rm exp}$ gets weaker. As a result, the lightest neutralino ($\tilde\chi_1^0$) can also have a relatively larger mass. 

%%%%%%%%%%%%%%%%%%%%%%%%%%%%%%%%%%
\section{Summary and Conclusion}
\label{sec:5}
%%%%%%%%%%%%%%%%%%%%%%%%%%%%%%%%%%

In the Minimal Supersymmetric Standard Model (MSSM), a dominantly Bino-like neutralino with a minimal mixing of (i) Higgsino and (ii) Higgsino-Wino components can effectively explain both the observed relic density of DM and the current data on the muon magnetic moment anomaly, denoted as $a_\mu$. Considering the anomaly in $a_\mu$ we additionally use the BMW and CMD3 results for SM prediction. In the $\tilde{B}_{\tilde H}$ scenario, a precise adjustment of the slepton mass is essential to facilitate effective slepton coannihilation processes to yield the DM abundance in the right ballpark. On the other hand, the $\tilde{B}_{\tilde{W}\tilde{H}}$ case involves chargino coannihilation in addition to the slepton coannihilations.
Although the acceptability of these scenarios as a potential candidate beyond the standard model is irrefutable, the direct and indirect searches of these models may remain challenging in specific parts of the parameter space. In particular, 
with compressed SUSY spectra, {\it e.g.} close proximity of $m_{\tilde{\chi}_1^0}$\,-\,$m_{\tilde{l}_{L,R}}$ or $m_{\tilde{\chi}_1^0}$\,-\,$m_{\tilde{\chi}_1^\pm}$, one finds it difficult to probe the scenarios at the LHC. Similarly, with minimal mixings with
Higgsinos, the DM direct detections may
not be quite promising, even with the
future generations of experiments.
On the other hand, 
the present-day annihilation strength of the DM in galactic halos turns out to be very weak, resulting in faint indirect search signals in the conventional DM halo profile. 

However, the situation, especially regarding indirect searches, can be improved drastically in the presence of a density spike around an SMBH ({\it e.g.} Sgr\,A$^*$) in the centre of the Milky Way. A sufficient density spike can provide an effective boost to the $\gamma$-ray flux that might be detectable in DM indirect search experiments. The spike profile is primarily characterized by three parameters, namely $\gamma_{\rm sp}$, $\gamma_{c}$ and $\rho_{\rm ann}$. An increase in either of the first two parameters results in a steeper DM spike around the SMBH and can lead to the desired boost to the DM annihilation rate in the GC.
In our analysis, we have kept $\gamma_c$ fixed at 1.2 ({\it i.e.} the maximum possible value) and considered $\gamma_{\rm sp}$ as a free parameter. On the other hand, $\rho_{\rm ann}$ depends on the DM mass and annihilation cross-section, thereby making $\rho(r)$, and consequently the $J$-factor, dependent on particle physics parameters.
Turning to the low energy data, along with the $a_\mu$ range, we undertake several constraints spanning from bounds on Higgs observables, SUSY particle searches at LHC, DM relic and SI-DD bounds to flavour constraints. We begin with portraying a few benchmark points, which indicate that it is indeed possible to reach the present sensitivity of the Fermi and HESS data for $\gamma$-ray searches associated with Sgr\,A$^*$ in the presence of an appropriate level of boost. We find that tuning $\gamma_{\rm sp}$ is sufficient to obtain the desired boost.
Subsequently, we use a few of the BMPs to unfold the role of some of the relevant parameters in the estimate of $\gamma$-ray energy flux.
Specifically, the parameter $\mu$, which predominantly determines the Higgsino fraction in both $\tilde{B}_{\tilde{H}}$ and $\tilde{B}_{\tilde{W}\tilde{H}}$ DM is found to play a crucial role in controlling the $\gamma$-ray flux for a particular DM mass. In the $\tilde{B}_{\tilde{W}\tilde{H}}$ scenario, the parameter $M_2$ also influences the $\gamma$-ray flux essentially in a similar manner.

Finally, we present the results of random scans showing the allowed ranges of DM mass for each category of $\tilde{B}_{\tilde{H}}$ and $\tilde{B}_{\tilde{W}\tilde{H}}$ for different choices of $\delta a_\mu$. We depict the minimum required value of $\gamma_{\rm sp}$ necessary to test the feasibility of each point based on indirect search data of Fermi and HESS. The preferred mass windows for $\tilde{B}_{\tilde{H}}$ and $\tilde{B}_{\tilde{W}\tilde{H}}$ are found to be $120\,{\rm GeV}\lesssim m_{\tilde{\chi}_1^0}\lesssim 360\,{\rm GeV}$ and $120\,{\rm GeV}\lesssim m_{\tilde{\chi}_1^0}\lesssim 530\,{\rm GeV}$ respectively. For a relaxed $\delta a_\mu$, such as CMD3, the $\gamma_{\rm sp}^{\rm min}$ recides on a relatively lower side. Overall, the requisite $\gamma_{\rm sp}^{\rm min}$ to eliminate the viable parameter space remains close to its adiabatic value, $\gamma_{\rm sp}\sim2.35$.

%%%%%%%%%%%%%%%%%%%%%%%%%%%%%%%%
\section*{Acknowledgement}
%%%%%%%%%%%%%%%%%%%%%%%%%%%%%%%%
The computations in this project were partially supported by SAMKHYA, the high-performance computing (HPC) facility provided by the Institute of Physics, Bhubaneswar (IOPB). The authors acknowledge S. Bisal, A. Chatterjee, B.T. Chiang, S. Rout, A. Tapadar and F. Varsi for valuable discussions.

%%%%%%%%%%%%%%%%%%%%%%%%%%%%%%%%%%%%%%%%%%%%%%%%%%%%%%%%%%%%%%%%%%%%%%%%%%%%%%
\bibliographystyle{JHEPCust.bst}
\bibliography{Bino_like_DM.bib}
%%%%%%%%%%%%%%%%%%%%%%%%%%%%%%%%%%%%%%%%%%%%%%%%%%%%%%%%%%%%%%%%%%%%%%%%%%%%%%

\end{document}